\documentclass[a4paper,11pt]{article}
\pdfoutput=1 
\usepackage{jheppub} 
\usepackage{amsmath,amsfonts,amssymb,mathrsfs,graphicx,color,longtable,bm,wasysym}
\usepackage{hyperref}
\usepackage{graphicx}
\usepackage{array}
\usepackage{color}
\usepackage[usenames,dvipsnames,table]{xcolor}
\usepackage{tikz}
\usepackage[utf8]{inputenc}
\usepackage{epsfig,relsize,xspace}
\usepackage[T1]{fontenc}
\usepackage{bbold}
\usepackage{float}
\usepackage{amsmath}
\usepackage{empheq}
\usepackage{booktabs}
\usepackage{color}
\usepackage[utf8]{inputenc}
\usepackage{xspace}
\usepackage{scalerel}
\usepackage[most]{tcolorbox}
\usepackage{multirow}
\usepackage{scalefnt}
\usepackage{bold-extra}
\usepackage[shortlabels]{enumitem}
\usepackage[tikz]{bclogo}
\usepackage{subcaption}
\usepackage{tikz-feynman}
\usepackage{cancel}
\usepackage{verbatim}
\usetikzlibrary{arrows,shapes}
\usepackage{afterpage}
\usepackage{graphicx,grffile}
\usepackage{slashed}
\usepackage{soul}
\usepackage{comment}

\usetikzlibrary{shapes.misc}

\tikzset{cross/.style={cross out, draw=black, line width=.4ex,minimum size=2*(#1-\pgflinewidth), inner sep=0pt, outer sep=0pt},
cross/.default={7pt}}

\allowdisplaybreaks
 
\definecolor{nicered}{rgb}{0.7,0.1,0.1}
\definecolor{nicegreen}{rgb}{0.1,0.5,0.1}
\definecolor{niceblue}{rgb}{0.0,0.1,0.7}
\hypersetup{colorlinks,citecolor=niceblue,linkcolor=niceblue,urlcolor=niceblue}
                                      
\def\bm#1{\mbox{\boldmath$#1$\unboldmath}}
\def \beq{\begin{equation}}
\def \eeq{\end{equation}}
\def \bea{\begin{eqnarray}}
\def \eea{\end{eqnarray}}

\begin{document}

\def\arraystretch{1.25}

\title{\boldmath Probing Higgs portals with matrix-element based kinematic discriminants in~$ZZ \to 4 \ell$ production}

\author[1]{Ulrich Haisch}

\author[1]{and Gabri{\"e}l Koole}

\affiliation[1]{Max Planck Institute for Physics, F{\"o}hringer Ring 6,  80805 M{\"u}nchen, Germany}

\emailAdd{haisch@mpp.mpg.de}
\emailAdd{koole@mpp.mpg.de}

\abstract{A Higgs portal in the form of the  operator~$|H|^2$  provides a minimal and theoretically  motivated link between the Standard Model~(SM) and  new physics. While~Higgs portals can  be constrained well by exotic Higgs decays if the~beyond-the-SM states are light, testing scenarios where these  particles are kinematically inaccessible is known to be challenging. We explore the sensitivity of  future hadron collider measurements  of~$ZZ \to 4 \ell$ production  in constraining   Higgs portal interactions. It is shown that  by using a matrix-element based kinematic discriminant the reach of the high-luminosity option of the Large Hadron Collider~(LHC) can be significantly enhanced compared to studies that are based on measurements of the four-lepton  invariant mass spectrum alone. We also analyse the  potential of the high-energy upgrade of the~LHC and a Future Circular Collider in constraining  new physics that couples to~$|H|^2$. The obtained constraints are compared to the limits one expects to find from other single-Higgs probes. In addition, we provide an independent analysis of the relevant Higgs portal effects in double-Higgs production. We find that the constraints obtained from our~$ZZ \to 4 \ell$ analysis turn out to be both competitive with and complementary to the projected limits obtained using other search techniques.
}

\maketitle

\section{Introduction}
\label{sec:introduction}

The discovery of a  new spin-0 state by ATLAS and CMS~\cite{ATLAS:2012yve,CMS:2012qbp} with approximately the properties of the Standard~Model~(SM) Higgs boson has opened up new avenues in the pursuit of physics beyond the SM~(BSM). In fact, there are both experimental and theoretical arguments that suggest that the Higgs boson may provide a window into BSM physics. Experimentally, the Higgs sector is far less explored and constrained compared to the gauge or fermionic sector of the SM~\cite{CMS-PAS-HIG-19-005,ATLAS-CONF-2020-027}, while theoretically the SM Higgs doublet~$H$ plays a special role because it allows to write down relevant and marginal  operators of the form~$|H|^2 \hspace{0.25mm} {\cal O}$ with~${\cal O}$ itself a gauge-invariant operator with a mass dimension of two or lower. 

The simplest and most studied case of such an operator is~${\cal O} = \phi^2$ where~$\phi$ is a real scalar  that is a singlet under the SM gauge group but odd under a~$ \mathbb{Z}_2$  symmetry~\cite{Silveira:1985rk,McDonald:1993ex,Burgess:2000yq,Patt:2006fw,Barger:2007im}. The corresponding interaction Lagrangian reads
\beq \label{eq:LHphi}
{\cal L}_{H\phi} = -c_\phi \hspace{0.5mm} |H|^2  \hspace{0.25mm} \phi^2 \,.
\eeq
Notice that  the~$ \mathbb{Z}_2$  symmetry acts on the real scalar field as~$\phi \to -\phi$, which guarantees the stability of~$\phi$  making it a suitable dark matter~(DM) candidate. See for instance~\cite{Arcadi:2019lkac,Lebedev:2021xey,Argyropoulos:2021sav} for recent reviews of the ensuing DM phenomenology. In particular, under the assumption that~$\phi$ is a relic of standard thermal freeze-out production 
DM direct detection experiments are known to foster stringent constraints on DM portals of the form (\ref{eq:LHphi}) --- see~for~example~\cite{Argyropoulos:2021sav} and references therein. In theories with a non-thermal cosmological history, a real scalar  $\phi$ can however be shown to be a viable DM candidate  for a wide range of Higgs portal realisations while evading existing experimental limits~\cite{Hardy:2018bph}. This opens up the possibility to probe~(\ref{eq:LHphi}) at high-energy colliders.  

Another motivation for the existence of sizeable Higgs portal couplings to~$|H|^2$ is provided by the  hierarchy problem of the Higgs-boson  mass. In fact, in models where the hierarchy problem is addressed by the addition of~$N_r$ real scalar  top partners~$\phi_i$ the relevant interaction Lagrangian can be written as~\cite{Curtin:2015bka}
\beq  \label{eq:LHstop}
{\cal L}_{H \phi_i} =- \frac{2 N_c}{N_r}  \, y_t^2 \hspace{0.5mm} |H|^2  \sum_{i=1}^{N_r}  \phi_i^2 \,, 
\eeq
where~$N_c = 3$ is the number of colours in QCD and~$y_t =\sqrt{2}  \hspace{0.25mm} m_t/v\simeq 0.94$ is the top-quark Yukawa coupling with~$m_t \simeq 163 \, {\rm GeV}$ the top-quark~$\overline{\rm MS}$ mass and~$v \simeq 246 \, {\rm GeV}$ the~Higgs vacuum expectation value. Well-known cases where~(\ref{eq:LHstop}) is a proxy for the resulting Higgs~portal interactions are  stops in the minimal supersymmetric SM~(MSSM)  and singlet scalar top partners in the hyperbolic Higgs~\cite{Cohen:2018mgv} or  tripled top model~\cite{Cheng:2018gvu}, if one assumes that these particles are approximately degenerate in mass. Notice that in such a case the interactions~(\ref{eq:LHphi}) and~(\ref{eq:LHstop}) are equivalent from the perspective of collider phenomenology if~$|c_\phi| = 2 N_c/\sqrt{N_r} \hspace{0.5mm} y_t^2$.  In  the case of the MSSM, the hyperbolic Higgs and the tripled top model where~$N_r = 12$, a~light Higgs boson is therefore natural if one effectively has a Higgs portal  of the form~(\ref{eq:LHphi}) with coupling strength~$|c_\phi| \leq \sqrt{3}\hspace{0.5mm} y_t^2 \simeq 1.5$. 

The level of difficulty to discover or to exclude Higgs portals of the form~(\ref{eq:LHphi}) and~(\ref{eq:LHstop}) at high-energy colliders depends mainly on the  mass~$m_\phi$ of the new states that couple to~$|H|^2$. While in the case of~$m_\phi < m_h/2 \simeq 62.5 \, {\rm GeV}$  the decays of the Higgs boson into invisible~\cite{Djouadi:2011aa,Mambrini:2011ik,Djouadi:2012zc,CMS:2018yfx,ATLAS-CONF-2020-052} or undetected~\cite{ATLAS-CONF-2020-027,Argyropoulos:2021sav} final states provide stringent constraints on the effective coupling strength of the Higgs portals,  obtaining relevant constraints above the kinematic threshold~$m_\phi > m_h/2$ turns out to be significantly more challenging. In~fact, only two categories of collider measurements  are known that provide sensitivity to Higgs portals above the kinematic threshold: firstly, pair-production of the new scalars in off-shell Higgs processes such as the vector-boson fusion~(VBF), the~$t \bar t h$ and the gluon-gluon-fusion~(ggF) channel~\cite{Glover:1988fe,Matsumoto:2010bh,Kanemura:2011nm,Kauer:2012hd,Chacko:2013lna,Endo:2014cca,Curtin:2014jma,Craig:2014lda,Ko:2016xwd,Buttazzo:2018qqp,Ruhdorfer:2019utl,Heisig:2019vcj,Englert:2020gcp,Garcia-Abenza:2020xkk,Haisch:2021ugv}, and secondly, studies of  the virtual effects that these particles produce when exchanged in loop diagrams that contribute to processes such as associated~$Zh$, double-Higgs  and~$gg \to h^\ast \to ZZ$ production~\cite{Englert:2013tya,Craig:2013xia,He:2016sqr,Kanemura:2016lkz,Goncalves:2017iub,Goncalves:2018pkt,Englert:2019eyl}. The existing analyses have considered a wide range of future high-energy hadron as well as lepton colliders, including the high-luminosity~(HL) and high-energy (HE) versions of the Large Hadron Collider~(LHC), a Future Circular Collider~(FCC), the International Linear Collider~(ILC), the Compact Linear Collider~(CLIC) and a muon collider.

In this article, we investigate  the sensitivity of future hadron collider measurements of off-shell Higgs production in the~$pp \to ZZ \to 4 \ell$ channel to  Higgs portal interactions such as~(\ref{eq:LHphi}) and  (\ref{eq:LHstop}). Compared to  earlier studies~\cite{Goncalves:2017iub,Goncalves:2018pkt,Englert:2020gcp,Caola:2013yja} that relied on the four-lepton invariant mass~($m_{4\ell}$) spectrum alone to separate signal from background, we instead employ a matrix-element~(ME) based kinematic discriminant in our work. Being~sensitive not only to~$m_{4 \ell}$ but also to another seven variables such as the invariant masses of the two opposite-sign lepton pairs (for details consult the articles~\cite{Gao:2010qx,Bolognesi:2012mm,Anderson:2013afp,Campbell:2013una}), ME-based discriminants fully exploit the event kinematics.  As in our recent study~\cite{Haisch:2021hvy}, we find that the use of a ME method leads to a significantly improved coverage of the BSM parameter space,~i.e.~$c_\phi$ and~$m_\phi$ in the case of~(\ref{eq:LHphi}), than a shape analysis of the~$m_{4 \ell}$ distribution.  Motivated by this finding, we  analyse in detail the  HL-LHC, HE-LHC and FCC potential of the proposed method in constraining  BSM physics that couples to the operator~$|H|^2$. 

Our work is structured as follows. In Section~\ref{sec:calculation} we briefly discuss the calculation of the loop corrections to~$pp \to ZZ \to 4 \ell$ production arising from~(\ref{eq:LHphi}). The aforementioned ME-based kinematic discriminant is introduced in Section~\ref{sec:ME} where we also discuss how higher-order QCD corrections are taken into account in our study.  The numerical analysis of the HL-LHC reach is performed in Section~\ref{sec:numHLLHC} and contains a comparison between the sensitivities obtained from  a shape analysis of the~$m_{4 \ell}$ spectrum and the proposed ME method. In~Section~\ref{sec:numHELHCFCC} we present our HE-LHC and FCC projections.   We~discuss our main results  in~Section~\ref{sec:conclusions}, comparing them to the limits one expects to obtain from other single- and double-Higgs probes, and provide a short outlook. A discussion of the impact that different assumptions on the systematic uncertainties in our  ME-based  search strategy have on the projected constraints is relegated to Appendix~\ref{app:errorstudy}, while further details of the relevant loop calculations and their implementation in the Monte Carlo~(MC) code for our double-Higgs analysis are given in Appendix~\ref{app:hhdetails}. 

\section{Higgs portal effects in~$\bm{gg \to h^\ast \to ZZ}$}
\label{sec:calculation}

\begin{figure}[t]
\centering
\vspace{.5cm}
\begin{tikzpicture}
         \begin{feynman}
        \vertex (a1)  {\(g\)};  
        \vertex[right=1.5cm of a1] (a2);
        \vertex[below=1.7cm of a1] (b1)  {\(g\)};
        \vertex[right=1.5cm of b1] (b2);
        \vertex[below=0.85cm of a2] (c1);
        \vertex[right=1.2cm of c1] (c2);
        \vertex[right=0.925cm of c2] (h2);
        \vertex[right=1.2cm of h2] (h3);
        \vertex[right=0.825cm of h3] (h4);
        \vertex[right=1cm of h4] (c3);
        \vertex[above=.8cm of c3] (c4) {\(Z\)};
        \vertex[below=.8cm of c3] (c5) {\(Z\)};
       \diagram* {
        (a1) -- [gluon, thick] (a2);
        (b1) -- [gluon, thick] (b2);
        (a2) -- [fermion, thick, edge label=\(t\)] (b2) -- [fermion, thick, edge label'=\(t\)] (c2) -- [fermion, thick, edge label'=\(t\)] (a2);
        (c2) -- [scalar, thick,  edge label=\(h\)] (h2);
        (h2) -- [scalar, half right, looseness=1.8, edge label'=\(\phi\), thick] (h3);
        (h2) -- [scalar, half left, looseness=1.8, edge label=\(\phi\), thick] (h3);
        (h3) -- [scalar, thick,  edge label=\(h\)] (h4);
        (h4) -- [boson, thick] (c4);
        (h4) -- [boson, thick] (c5);
         };
         \filldraw[color=black, fill=black] (a2) circle (0.04);
         \filldraw[color=black, fill=black] (b2) circle (0.04);
         \filldraw[color=black, fill=black] (c2) circle (0.04);
         \filldraw[color=black, fill=black] (h4) circle (0.04);
        \node[square dot,fill=black,scale=1.5] (d) at (h2){};
        \node[square dot,fill=black,scale=1.5] (d) at (h3){};
      \end{feynman}
\end{tikzpicture}
\vspace{.5cm}
\caption{\label{fig:propcorr} Example of a one-loop correction to~$gg \to h^* \to ZZ$ production with insertions of the Higgs portal operator~(\ref{eq:LHphi}) indicated by the black boxes. Also diagrams with tadpoles or counterterms contribute but are not shown explicitly. Consult the main text for further details.}
\end{figure}
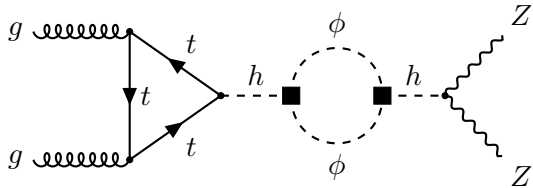

At the one-loop level the~$gg \to h^\ast \to ZZ$ process receives contributions from Feynman graphs such as the one displayed in Figure~\ref{fig:propcorr} that contains a modified Higgs propagator with insertions of the  Higgs portal operator~(\ref{eq:LHphi}). The corresponding renormalised contribution to the self-energy of the Higgs takes the form
\begin{align}\label{eq:renprop}
\begin{split}
 \hat{\Sigma}(\hat s) =  \Sigma(\hat s) +  \left (\hat s - m_h^2 \right ) \delta Z_h -  \delta m_h^2  \ ,
\end{split}
\end{align}
where the bare Higgs self-energy, the  one-loop corrections to the Higgs wave function and the mass counterterm in the on-shell scheme are given by the following expressions
\begin{align} \label{eq:proping}
\begin{split}
\Sigma(\hat s) &= \frac{1}{(4 \pi)^2} \hspace{0.25mm}  \Big [ c_\phi \hspace{0.25mm} A_0 (m_\phi^2) + 2 \hspace{0.25mm}  v^2  \hspace{0.25mm} |c_\phi |^2 \hspace{0.25mm}  B_0 \! \left (\hat s,m_\phi^2,m_\phi^2 \right )  \Big ] \,, \\[2mm]
\delta Z_h &=  -\frac{2 \hspace{0.25mm}  v^2 \hspace{0.25mm} |c_\phi |^2}{(4 \pi)^2}   \! \left . \frac{d}{d \hat s}  B_0 \! \left (\hat s,m_\phi^2,m_\phi^2 \right ) \right |_{\hat s = m_h^2} \,,  \\[2mm]
\delta m^2_h &=  \frac{1}{(4 \pi)^2} \hspace{0.25mm}  \Big [ c_\phi \hspace{0.25mm} A_0 (m_\phi^2) + 2 \hspace{0.25mm}  v^2  \hspace{0.25mm} |c_\phi |^2 \hspace{0.25mm}  B_0 \! \left (m_h^2,m_\phi^2,m_\phi^2 \right )  \Big ]  \,.
\end{split}
\end{align}
Here~$\hat s = p^2$ with~$p$ the external four-momentum entering the Higgs propagator and the~$A_0$ and~$B_0$ functions are  one- and two-point Passarino-Veltman scalar integrals defined as in~\cite{Hahn:1998yk,Hahn:2016ebn}. The expression in (\ref{eq:proping}) can be easily generalised  to other Higgs portals of the form~(\ref{eq:LHphi}). For~instance,  in the case of~${\cal L}_{H \Phi} =  -c_\Phi \hspace{0.5mm} |H|^2  \hspace{0.25mm} |\Phi|^2$ with~$\Phi$ a complex scalar field one just has to make the substitutions~$c_\phi \to c_\Phi/\sqrt{2}$ and~$m_\phi \to m_\Phi$. 

Notice that the contribution to the Higgs wave-function renormalisation constant~$\delta Z_h$ coming from the propagator corrections  exactly cancels against those of the vertices when combined to obtain  the full BSM contribution to the off-shell~$gg \to h^\ast \to Z Z$ amplitude. Similarly, the tadpole contribution proportional to~$A_0 (m_\phi^2)~$ also cancels in the difference~$\Sigma(\hat s) -  \delta m_h^2$. In contrast,  the Higgs wave-function renormalisation constant $\delta Z_h$ does not  drop out in the on-shell Higgs signal strengths~$\mu_{i}^f$ for production in channel~$i$ and decay in channel~$f$. In terms of the  inclusive Higgs production cross sections~$\sigma_i$ and the Higgs branching ratios~${\rm BR}_f$, these quantities take the form 
\beq \label{eq:signalstrengths}
\mu_{i}^f = \frac{\sigma_i}{\sigma_i^{\rm SM}}  \frac{{\rm BR}_f}{{\rm BR}^{\rm SM}_f} = 1 + \delta Z_h \,, 
\eeq 
i.e.~they receive a universal correction proportional to the Higgs wave-function renormalisation constant as given in~(\ref{eq:proping}). This feature  allows to set indirect constraints on Higgs portal models by precision measurements of  Higgs properties~\cite{Craig:2013xia}, which will be discussed in Section~\ref{sec:conclusions}.

\section{ME-based kinematic discriminant}
\label{sec:ME} 

The Higgs propagator corrections~(\ref{eq:renprop}) and the relevant vertex counterterms have been implemented into version 8.0 of the event generator {\tt MCFM}~\cite{Boughezal:2016wmq} to obtain kinematic distributions  for~$pp \to ZZ \to 4 \ell$ such as the~$m_{4 \ell}$ spectrum. In addition, our MC code is also able to calculate the following ME-based kinematic discriminant~\cite{ATLAS:2015cuo,ATLAS:2018jym,ATLAS:2019qet} 
\beq \label{eq:kindiscr}
D_S = \log_{10} \Bigg( \frac{P_h}{P_{gg} + c \cdot P_{q\bar{q}}} \Bigg) \, .
\eeq
Here~$P_h$ denotes the squared ME  for the~$gg \to h^\ast \to Z Z \to 4 \ell$ process,~$P_{gg}$ is the squared ME  for all~$gg$-initiated channels (including the Higgs channel, the continuum background and their interference) and~$P_{q\bar{q}}$ is the squared ME  for the~$q\bar{q} \to Z Z \to 4\ell$ process. Like~in~\cite{ATLAS:2015cuo,ATLAS:2018jym,ATLAS:2019qet} the constant~$c$ is set to 0.1 to balance the~$q\bar{q}$- and~$gg$-initiated contributions.
We stress that in the SM more than 99\% of the~$pp \to Z Z \to 4 \ell$  cross section falls into the range of~$-4.5 < D_S < 0.5$~\cite{ATLAS:2015cuo}. For BSM models that predict events with~$D_S < -4.5$ or~$D_S > 0.5$ the variable~$D_S$ therefore presents a null test.

Currently, calculations of higher-order QCD corrections to four-lepton production  via $q \bar q$~annihilation include the full next-to-next-to-leading order (NNLO) corrections and top-quark mass effects~\cite{Cascioli:2014yka,Grazzini:2015hta,Heinrich:2017bvg,Kallweit:2018nyv}. Next-to-leading order (NLO) corrections to the loop-induced $gg$~channel have been computed by now as well \cite{Caola:2015psa,Caola:2016trd,Grazzini:2018owa,Grazzini:2021iae}, while for inclusive Higgs production  the precision has been pushed to the the next-to-next-to-next-to-leading order~(N$^3$LO) in the heavy top-quark limit~\cite{Anastasiou:2015vya}. As the $gg \to ZZ$ process starts contributing only at $\mathcal{O}(\alpha_s^2)$, it is part of the NNLO QCD corrections to $ZZ$ production and NLO corrections to this channel formally contribute at N$^3$LO. Lastly, NLO electroweak (EW) corrections could in principle play an important role as well. Within the SM they were combined with~NNLO~QCD effects for $ZZ$ production in the work~\cite{Grazzini:2019jkl}. However, it has been shown in the paper~\cite{Maltoni:2017ims} that including NLO EW effects in the SM has only a very minor effect on the sensitivity of indirect single-Higgs analyses to modifications of the trilinear Higgs coupling. We expect a similar pattern to arise in the context of the Higgs portal models studied here. A dedicated simulation of four-lepton events including both higher-order QCD as well as EW corrections both in and beyond the SM, consistently matched to a parton~shower and including detector effects is clearly beyond the scope of the present article and therefore left for future work. 

In order to include higher-order QCD corrections in our~$pp \to ZZ \to 4 \ell$ analysis, we proceed as in our recent publication~\cite{Haisch:2021hvy}. For the two relevant  production channels we calculate the so-called~$K$-factor defined as the ratio between the fiducial cross section at a given order in QCD and  the corresponding leading order~(LO) QCD prediction. In the case of the~$gg$-initiated contribution we utilise the results of~\cite{Buonocore:2021fnj}. The ratio between the NLO and LO~ggF predictions turns out to be essentially flat  in~$m_{4 \ell}$ and by~averaging we find~$K^{\rm NLO}_{gg}  = 1.83$. This number agrees with the~$K$-factors reported in~\cite{Caola:2015psa,Grazzini:2018owa,Grazzini:2021iae}. In~the case of the~$q \bar q$-initiated contribution we use the NNLO results obtained in~\cite{Grazzini:2018owa}. The relevant~$K$-factor again turns out to be basically flat in~$m_{4\ell}$ with a central value of~$K^{\rm NNLO}_{q\bar q}  = 1.55$. This~finding is in accordance with~\cite{Cascioli:2014yka}. The quoted~$K$-factors are then used to obtain a QCD-improved prediction  for the~$pp \to ZZ \to 4 \ell$ cross section differential in the variable~$O$ as follows:
\beq \label{eq:procedure}
\frac{d \sigma_{pp} }{d O}  = K^{\rm NLO}_{gg}  \left ( \frac{d \sigma_{gg}}{d O} \right )_{\rm LO} +  K^{\rm NNLO}_{q \bar q}  \left ( \frac{d \sigma_{q \bar q}}{d O} \right )_{\rm LO} \,.
\eeq
Notice that~(\ref{eq:procedure}) is accurate in the case of the~$m_{4 \ell}$ spectrum. For~the~$D_S$ distribution one observes~\cite{Haisch:2021hvy} a close to flat~$K$-factor of around~$1.6$ between the LO and the improved prediction~(\ref{eq:procedure}). It is furthermore found that the inclusion of higher-order QCD corrections reduces the scale uncertainties by a factor of about 3 from~$(7 - 8)\%$ to~$(2 - 3)\%$. The fact that the central value of the improved~$D_S$ spectrum lies  outside the~LO uncertainty bands demonstrates that the scale variations of~(\ref{eq:procedure}) do not provide a reliable way to estimate the size of higher-order QCD effects. In view of this and given  that the  discriminant~$D_S$ as defined in~(\ref{eq:kindiscr}) is only LO accurate, we will make different assumptions on the systematic uncertainties entering our ME-based search strategy, a point we will discuss in more detail in our numerical analyses presented in Sections~\ref{sec:numHLLHC} and~\ref{sec:numHELHCFCC} as well as in~Appendix~\ref{app:errorstudy}. A~similar approach is also used in the projections~\cite{ATL-PHYS-PUB-2015-024,CMS-PAS-FTR-18-011} that estimate the HL-LHC reach in constraining off-shell Higgs boson production and the Higgs boson total width in~$pp \to ZZ \to 4 \ell$. 
 
\section{HL-LHC analysis}
\label{sec:numHLLHC} 

In our~$pp \to ZZ \to 4 \ell$ analysis we consider the window~$140 \, {\rm GeV} <m_{4 \ell}<  600 \, {\rm GeV}$ of four-lepton invariant masses.   The~charged leptons are required to be  in the pseudorapidity range~$|\eta_\ell| < 2.5$ and the lepton with the highest transverse momentum~($p_T$)  must satisfy~$p_{T,\ell_1} > 20 \, {\rm GeV}$ while the second, third and fourth  lepton in~$p_T$ order is required to obey~$p_{T,\ell_2} > 15 \, {\rm GeV}$,~$p_{T,\ell_3} > 10 \, {\rm GeV}$ and~$p_{T,\ell_4} > 6 \, {\rm GeV}$, respectively.  The lepton pair with the mass closest to the~$Z$-boson mass is referred to as the leading dilepton pair and its invariant mass is required to be within~$50 \, {\rm GeV} < m_{12} < 106 \, {\rm GeV}$, while the  subleading lepton pair must be in the range of~$50 \, {\rm GeV} < m_{34} < 115 \, {\rm GeV}$. Notice that the ATLAS and CMS analyses~\cite{CMS:2014quz,ATLAS:2015cuo,ATL-PHYS-PUB-2015-024,CMS-PAS-FTR-18-011,ATLAS:2018jym,CMS:2019ekd,ATLAS:2019qet} employ similar cuts.  We assume a detection efficiency of 99\% (95\%) for muons (electrons) that satisfy the event selections. These efficiencies correspond to those reported  in the latest ATLAS analysis of off-shell Higgs production~\cite{ATLAS:2019qet}. As input parameters we use~$G_F = 1/(\sqrt{2} \hspace{0.25mm} v^2) = 1.16639 \hspace{0.25mm}\cdot \hspace{0.25mm}10^{-5}\,  {\rm GeV}^{-2}$,~$m_Z = 91.1876 \ \text{GeV}$,~$m_h = 125 \,  {\rm GeV}$ and~$m_t = 173 \, {\rm GeV}$.  We employ {\tt NNPDF40\_nlo\_as\_01180} parton distribution functions~(PDFs)~\cite{Ball:2021leu} with the renormalisation and factorisation scales~$\mu_R$ and~$\mu_F$ set to~$m_{4\ell}$ on an event-by-event basis. Both the different-flavour~$e^+ e^- \mu^+ \mu^-$ and the same-flavour~$2e^+ 2e^-$ and~$2\mu^+ 2 \mu^-$ decay channels of the two~$Z$ bosons are included throughout our work.

\begin{figure}[!t]
\begin{center} 
\includegraphics[width=0.475\textwidth]{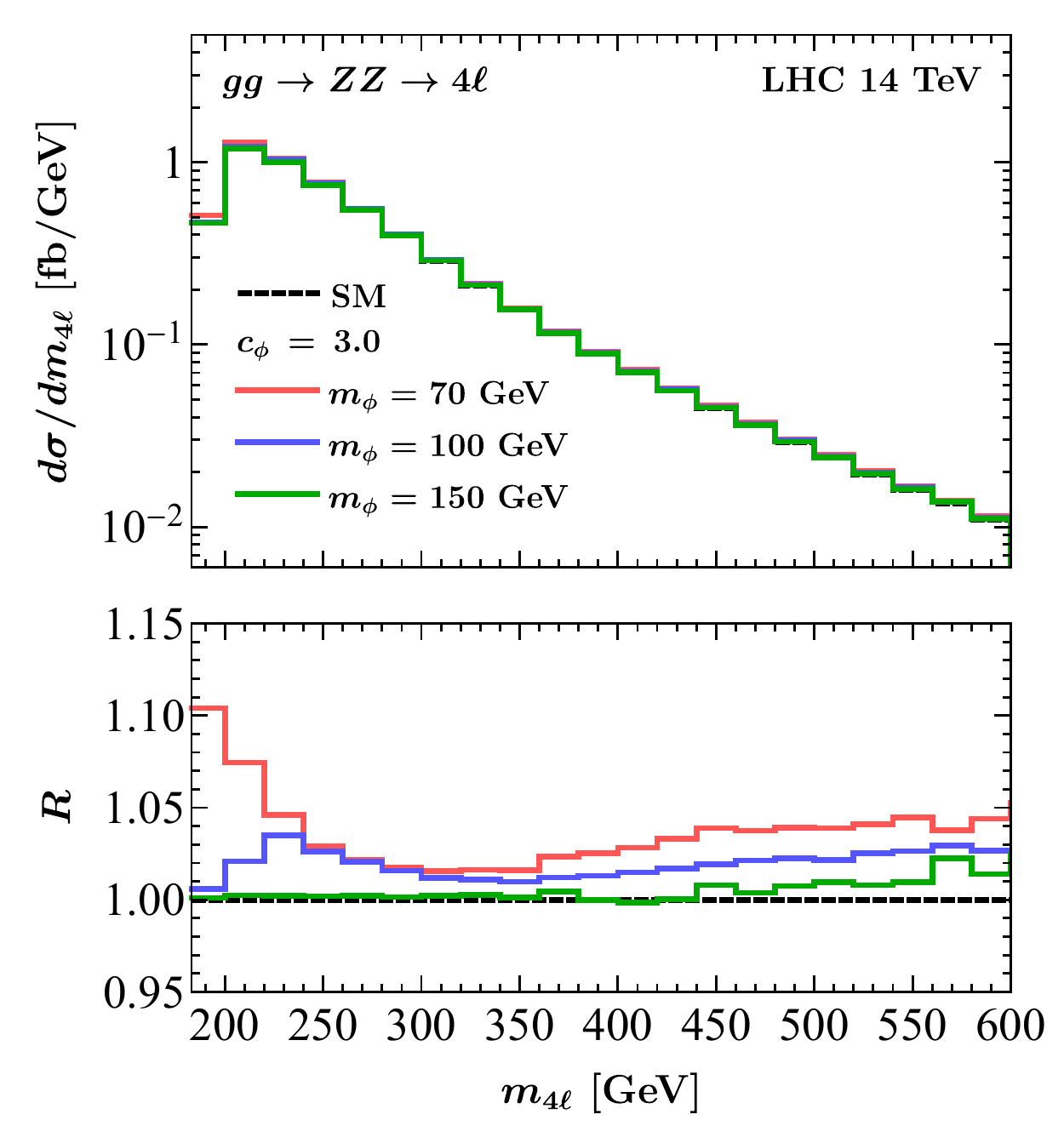} \quad 
\includegraphics[width=0.475\textwidth]{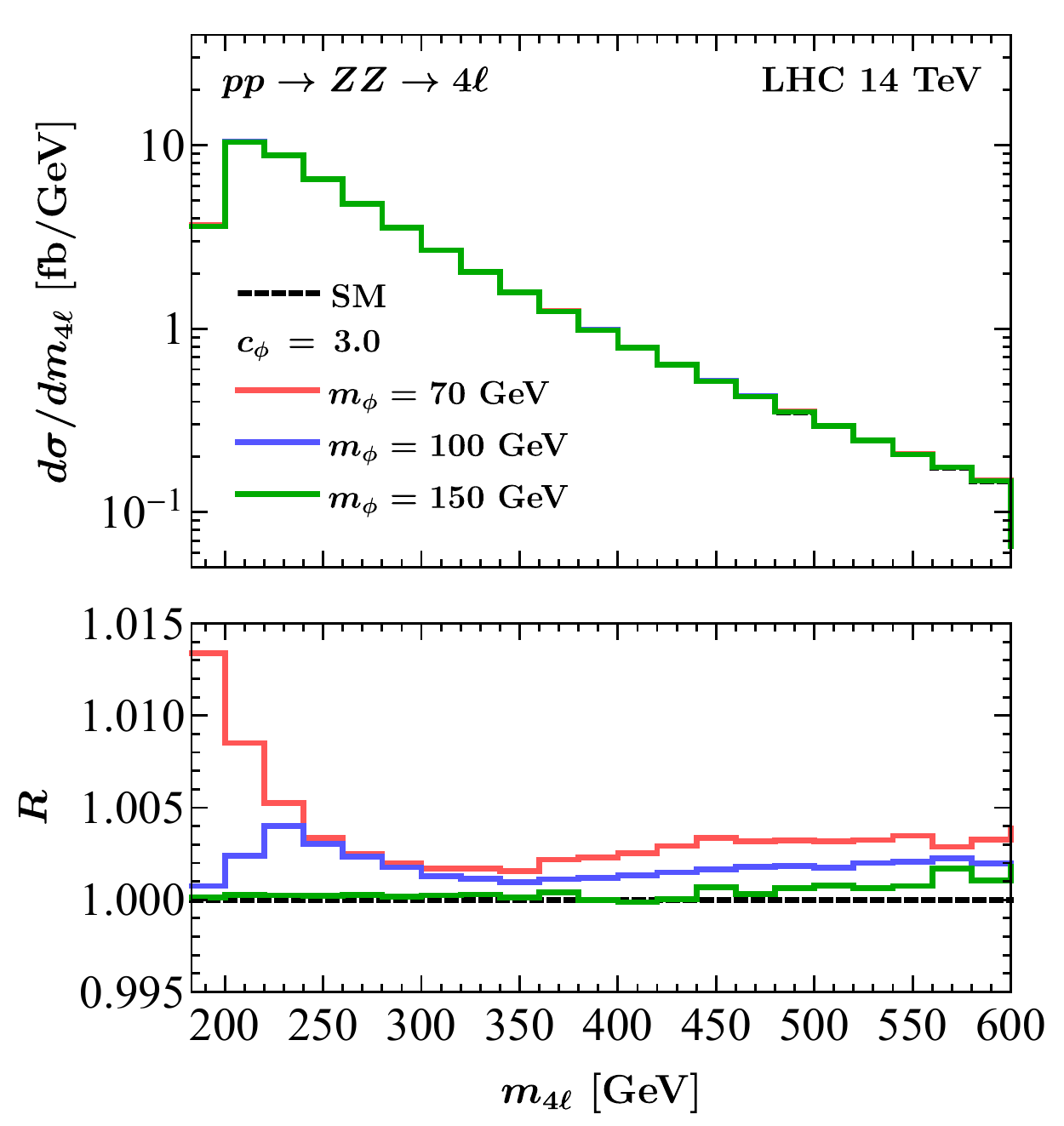}
\vspace{2mm} 
\caption{\label{fig:m4lplotHLLHC}$m_{4 \ell}$~spectra in the~SM~(dashed black) as well as  for three  Higgs portal model scenarios~(\ref{eq:LHphi}) assuming~$c_\phi =3$ and~$m_\phi = 70 \, {\rm GeV}$~(solid red), $m_\phi = 100 \, {\rm GeV}$~(solid blue) and $m_\phi = 150 \, {\rm GeV}$~(solid green). The left (right) plot shows results for~$gg \to ZZ \to 4 \ell$ ($pp \to ZZ \to 4 \ell$) production.  All~distributions correspond to QCD-improved predictions and LHC collisions at a centre-of-mass energy of~$\sqrt{s} = 14 \, {\rm TeV}$. The~lower panels depict the ratios between the BSM distributions and the corresponding SM predictions. }
\end{center}
\end{figure}
 
In Figure~\ref{fig:m4lplotHLLHC} we show our predictions for the~$m_{4 \ell}$ distributions in the~SM~(dashed~black) and  three  Higgs portal models~(\ref{eq:LHphi}).  The displayed BSM benchmarks correspond to  scalar masses of~$m_\phi = 70 \, {\rm GeV}$~(solid red), $m_\phi = 100 \, {\rm GeV}$~(solid blue) and $m_\phi = 150 \, {\rm GeV}$~(solid green) assuming in all cases a coupling strength of~$c_\phi = 3$.  Notice that the chosen value of $c_\phi$ is  safely below the limit $\left | c_\phi \right | < 4 \pi$ following from perturbative tree-level unitarity (see for instance~\cite{Englert:2020gcp}). In the left panel the QCD-improved predictions for~$gg  \to Z Z \to 4 \ell$  production  including the Higgs signal, the continuum background and their interference are given.  Two features of the shown BSM spectra deserve a further discussion. First, one observes peak-like structures in the distributions slightly above the threshold~$m_{4 \ell} = 2 m_\phi$ of two-scalar production. Second, both spectra show an enhancement  at large~$m_{4 \ell}$ because in the limit of partonic centre-of-mass energies~$\hat s \to \infty$ the correction simplifies to~$\Sigma (\hat s) - \delta m_h^2 \simeq - v^2 |c_\phi|^2 /( 8 \pi^2) \ln \left ( \hat s/m_h^2 \right )$. This behaviour is easily derived from~(\ref{eq:proping}). Notice furthermore that the~$gg \to h^\ast \to ZZ \to 4 \ell$ amplitudes interfere destructively with the~$gg \to ZZ \to 4 \ell$ matrix elements so that the overall sign of the correction~$\Sigma (\hat s) - \delta m_h^2$ is effectively flipped.  One also sees that for the three chosen sets of Higgs portal parameters the relative corrections in the spectra amount to less than~$15\%$  over the whole range of~$m_{4\ell}$ values of interest. The same features are also observed in the right panel of Figure~\ref{fig:m4lplotHLLHC} which shows the corresponding predictions for~$pp \to ZZ \to 4 \ell$ production. Notice that in this case the relative modification are smaller by a factor of roughly 10 than for~$gg \to ZZ \to 4 \ell$ due to the addition of the~$q \bar q \to Z Z \to 4 \ell$ channel which receives no BSM correction.  

\begin{figure}[!t]
\begin{center}
\includegraphics[width=0.475\textwidth]{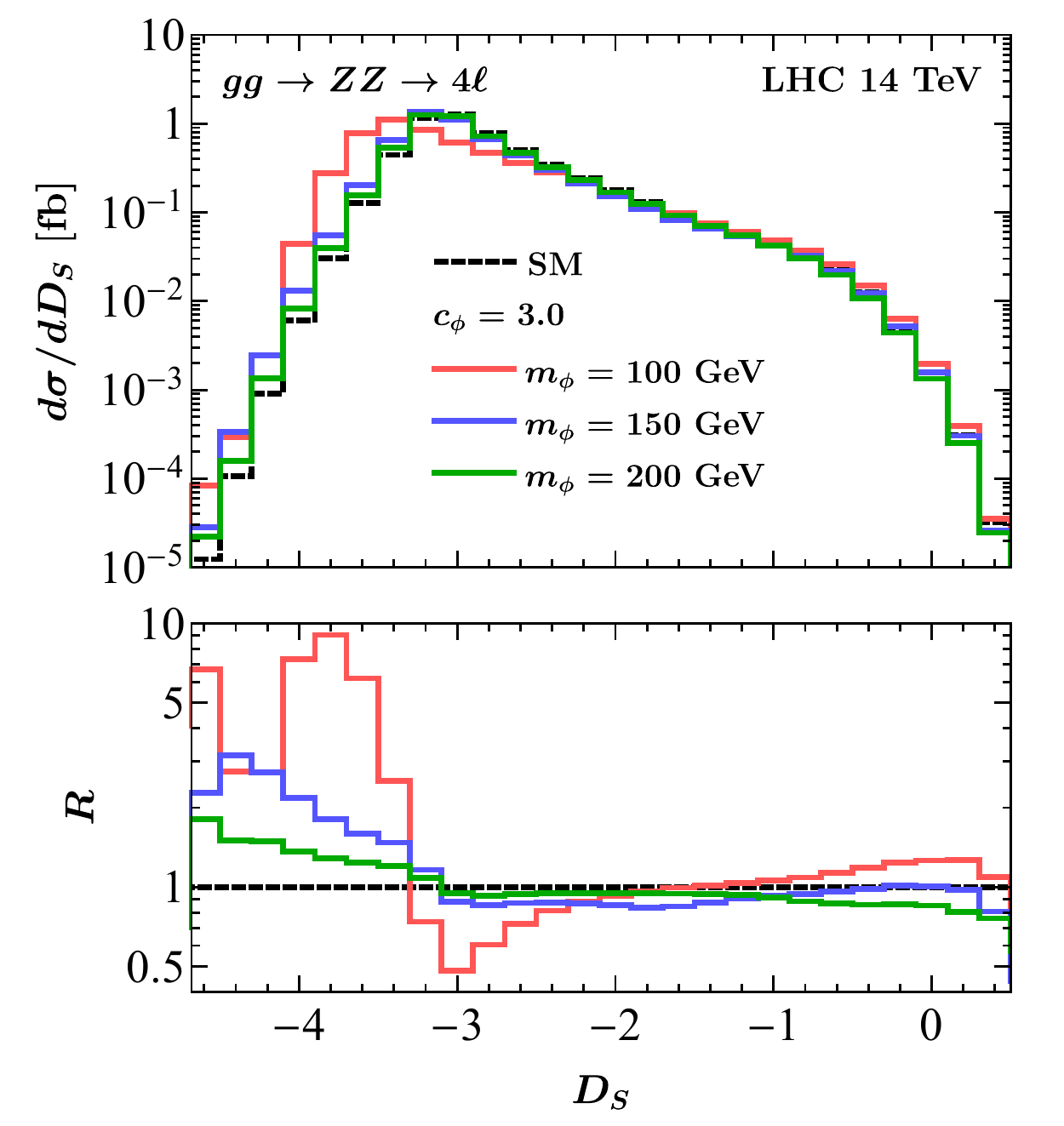} \quad 
\includegraphics[width=0.475\textwidth]{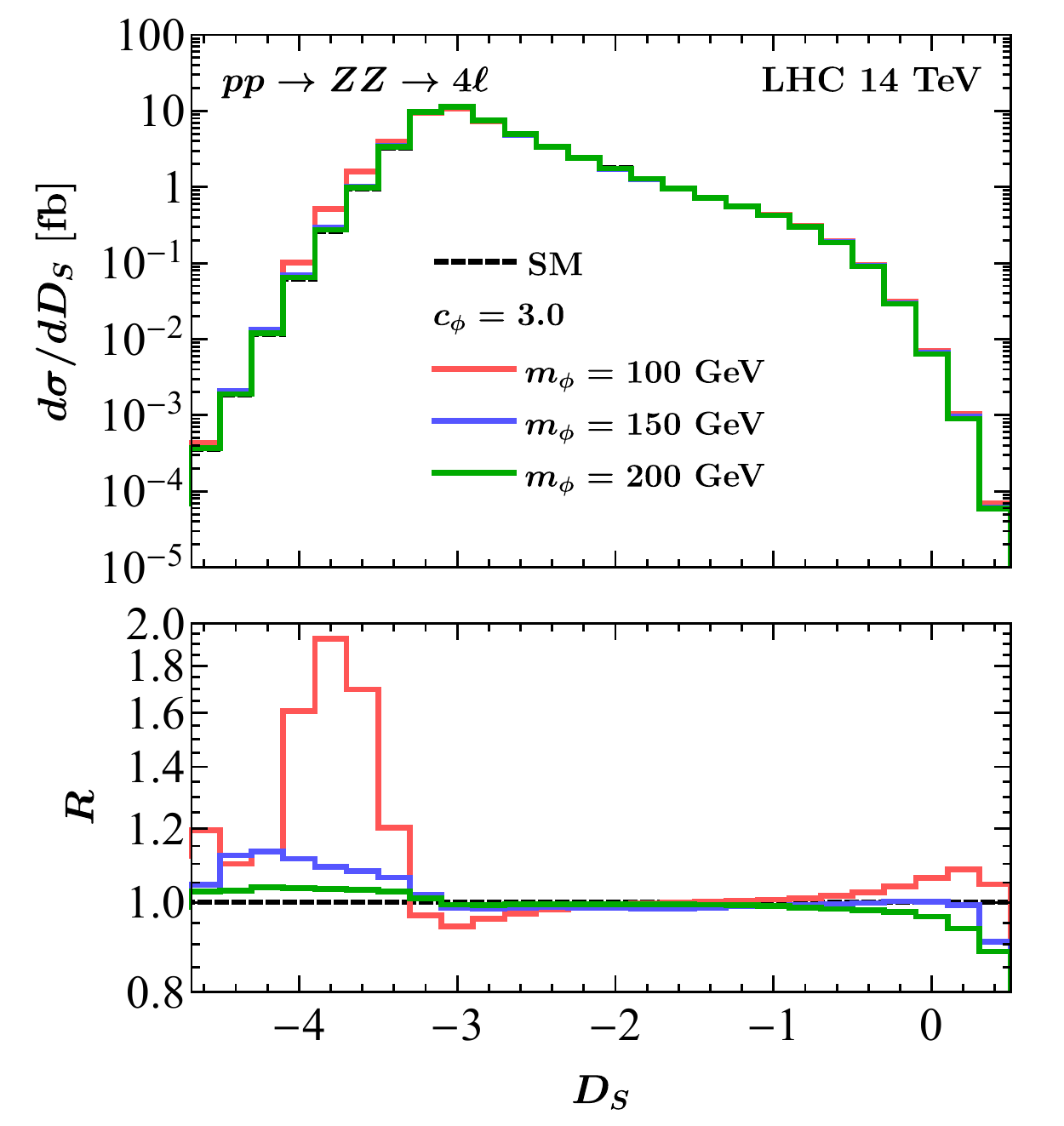}
\vspace{2mm} 
\caption{\label{fig:DSplotHLLHC} As Figure~\ref{fig:m4lplotHLLHC} but for the QCD-improved ME-based discriminant~$D_S$ as defined in~(\ref{eq:kindiscr}) and~(\ref{eq:procedure}). Furthermore, instead of $m_\phi = 70 \, {\rm GeV}$ a mass of $m_\phi = 200 \, {\rm GeV}$~is employed. For additional explanations see main text.}
\end{center}
\end{figure}

To illustrate the discriminating power of the ME-based kinematic variable introduced in~(\ref{eq:kindiscr}) we present in Figure~\ref{fig:DSplotHLLHC} the results for the~$D_S$ spectra in the SM and beyond. The shown predictions have been obtained by means of~(\ref{eq:procedure}) and the choices for the Higgs portal model parameters are those from before, apart from $m_\phi = 70 \, {\rm GeV}$  which is replaced by $m_\phi = 200 \, {\rm GeV}$. One observes that compared to the SM spectrum the BSM distributions are shifted to lower values of~$D_S$.  This is a simple consequence of the fact that the correction~$\Sigma (\hat s) - \delta m_h^2$ tends to reduce the~$gg \to h^\ast \to Z Z \to 4 \ell$ amplitude and thus~$P_h$ in~(\ref{eq:kindiscr}). As a result of the sharp cut-off of the SM distribution at~$D_S \simeq -3.5$, the relative BSM effects in the~$D_S$ spectra for~$gg \to Z Z \to 4 \ell$ turn out to be large, easily exceeding~$100 \%$ for the chosen benchmark values of~$c_\phi$ and~$m_\phi$. As illustrated in the right panel of Figure~\ref{fig:DSplotHLLHC}, adding the~$q \bar q \to ZZ \to 4 \ell$ channel to the predictions for the~$D_S$ distributions notably reduces the relative size of the Higgs portal corrections. Still assuming~$c_\phi =3$, the BSM effects reach the level of around 200\%, 10\% and 5\% in the case of~$m_{\phi} = 100 \, {\rm GeV}$, $m_{\phi} = 150 \, {\rm GeV}$ and   $m_{\phi} = 200 \, {\rm GeV}$, respectively. 

By comparing the relative modifications in the right panels of Figures~\ref{fig:m4lplotHLLHC} and~\ref{fig:DSplotHLLHC} it should be already  clear that the four-lepton invariant mass~$m_{4 \ell}$ has a much weaker discriminating power than the variable~$D_S$ in constraining interactions of the form~(\ref{eq:LHphi}). In order to make this statement quantitative we perform a shape  analysis of both the~$m_{4 \ell}$ and~$D_S$ spectrum following the method outlined in our earlier work~\cite{Haisch:2021hvy}. Specifically, the significance~$Z_{i}$ is calculated as a Poisson ratio of likelihoods modified to incorporate systematic uncertainties on the background using the Asimov approximation~\cite{Cowan:2010js}:
\beq \label{eq:Zi}
Z_{i} = \left \{ 2 \left [ \left (s_i + b_ i \right ) \ln \left [ \frac{\left (s_i + b_i \right) \left (b_i + \sigma_{b_i}^2 \right )}{b_i^2 + \left (s_i + b_i \right ) \sigma_{b_i}^2} \right ] - \frac{b_i^2}{\sigma_{b_i}^2}  \ln \left( 1+ \frac{s_i \hspace{0.25mm} \sigma_{b_i}^2}{b_i \hspace{0.25mm} (b_i + \sigma_{b_i}^2)} \right ) \right ]\ \right \}^{1/2} \,.
\eeq
Here~$s_i$~($b_i$) represents the expected number of signal (background) events in  bin~$i$ of the~$m_{4 \ell}$ or~$D_S$ spectrum and~$\sigma_{b_i}$ denotes the standard deviation that characterises the systematic uncertainties of the associated  background in that bin. To set bounds on~$c_{\phi}$ as a function of~$m_{\phi}$ we assume that the central values of a future measurements of the two  relevant distributions will line up  with the SM predictions. We hence employ 
\beq \label{eq:sbsigma}
s_i = N_i (c_\phi) - N_i (0) \,, \qquad b_i = N_i (0) \,, \qquad \sigma_{b_i} =  \Delta_i N_i (0)  \,.
\eeq 
The total significance~$Z$ is obtained by adding the individual~$Z_{i}$ values in quadrature. Parameter regions with a total significance of~$Z > \sqrt{2} \hspace{0.5mm} {\rm erf}^{-1} \left ( {\rm CL} \right )~$ are said to be excluded at a given confidence level CL. Here~${\rm erf}^{-1} (z)$ denotes the inverse error function. In our shape analyses, we consider 23 bins of size of~$20 \, {\rm GeV}$  with four-lepton invariant masses in the range~$140 \, {\rm GeV} < m_{4\ell}  < 600 \, {\rm GeV}$ and 27 bins of equal size of 0.2 that cover the range~$-4.9 < D_S  <0.5$ in the case of~$m_{4 \ell}$ and~$D_S$, respectively. 

A crucial ingredient in our analysis will turn out to be the systematic uncertainties~$\sigma_{b_i}$ on the background as parametrised by the parameters~$ \Delta_i$  in~(\ref{eq:sbsigma}).  In the case of the HL-LHC shape fits, we  will employ the two different choices~$\Delta_i = \Delta = 8\%$ and~$\Delta_i =  \Delta = 4\%$  of bin-independent systematic uncertainties. These choices can be motivated by recalling that the systematic uncertainties that ATLAS quotes in the HL-LHC study~\cite{ATL-PHYS-PUB-2018-054} for the on-shell~$gg \to h \to ZZ$ signal strength amount to~$5.0\%$ and~$3.9\%$ in the baseline scenario S1 and S2 for the expected total  systematic uncertainties. The corresponding systematic uncertainties quoted in the CMS work~\cite{CMS-PAS-FTR-18-011} are~$7.3\%$ and~$4.1\%$. Since the dominant Higgs portal corrections in~$D_S$ are associated to kinematic configurations with~$m_{4 \ell}$ around~$2 m_\phi$, we believe that for not too heavy~$\phi$,   theoretical predictions of the~$D_S$ spectra will reach an accuracy that is very similar to the systematics that is expected to be achievable at the HL-LHC in the case of on-shell~$gg \to h  \to ZZ$ production. Notice that the BSM effects in the~$m_{4 \ell}$ spectrum also receive important corrections in the region~$m_{4 \ell} > 2 m_t$ as can be seen from the plots in Figure~\ref{fig:m4lplotHLLHC}.  Given the limitations (cf.~\cite{Amoroso:2020lgh,Alioli:2021wpn,Buonocore:2021fnj}) of the state-of-the-art SM predictions of~$pp \to ZZ$ production for kinematic configurations above the two top-quark threshold,  achieving the assumed systematic uncertainties  of~$\Delta = 8\%$ and~$\Delta = 4\%$  is certainly more challenging in the case of the~$m_{4 \ell}$ distribution. The steady progress of perturbative QCD calculations, in particular the exact evaluations of the two-loop on-shell amplitudes for~$gg \to ZZ$ involving top quarks~\cite{Agarwal:2020dye,Bronnum-Hansen:2021olh} makes us, however, confident that systematic uncertainties in the ballpark of~10\% or below are  attainable till~$3 \, {\rm ab}^{-1}$ of data are collected at the HL-LHC. 

\begin{figure}[!t]
\begin{center}
\includegraphics[width=0.575\textwidth]{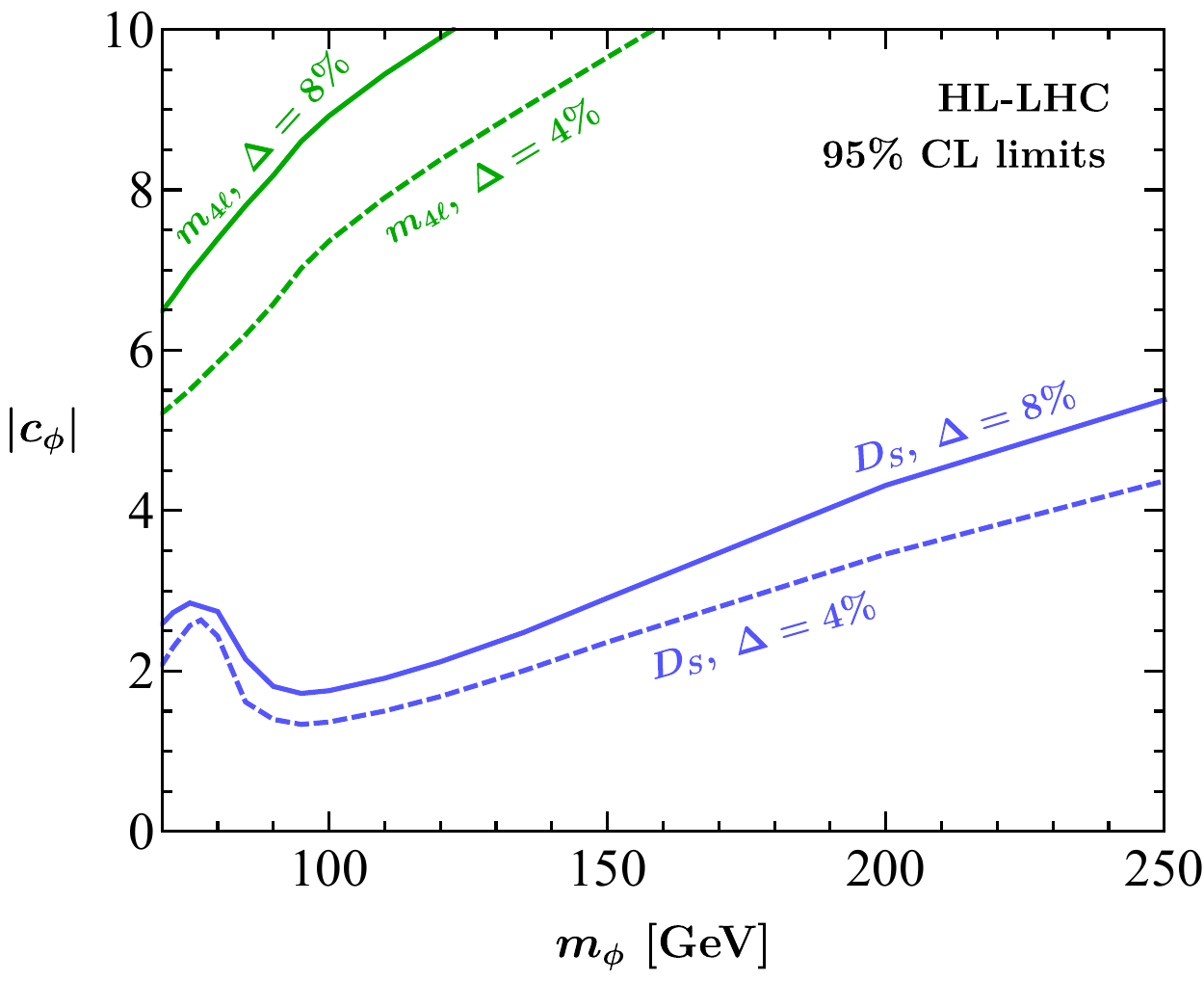} 
\vspace{2mm} 
\caption{\label{fig:HLLHCfits} 95\%~CL limits on~$|c_\phi|$  as a function of~$m_\phi$ derived from the binned-likelihood analysis of the~$m_{4 \ell}$~(green lines) and~the~$D_S$~(blue lines) spectrum at the HL-LHC. The~solid (dashed)  curves are obtained assuming a systematic uncertainty of~$\Delta = 8\%$~($\Delta = 4\%$). See~main text for additional details.   }
\end{center}
\end{figure}

The plot in  Figure~\ref{fig:HLLHCfits} displays  the results of our binned-likelihood analysis when applied to  the~$m_{4 \ell}$~(green lines) and~the~$D_S$~(blue lines) distribution. Given the strong constraints on~$c_\phi$ from on-shell Higgs boson decays into invisible~\cite{Djouadi:2011aa,Mambrini:2011ik,Djouadi:2012zc,CMS:2018yfx,ATLAS-CONF-2020-052} or undetected~\cite{ATLAS-CONF-2020-027,Argyropoulos:2021sav} final states, we only consider~$m_\phi$ values  above the Higgs threshold at~$m_h/2$.  The shown  95\%~CL limits  correspond to  our  HL-LHC projections assuming the full expected integrated luminosity of~$3 \, {\rm ab}^{-1}$ at~$\sqrt{s} = 14 \, {\rm TeV}$. The solid (dashed) exclusion lines have been obtained for a systematic uncertainty of~$\Delta = 8\%$~($\Delta = 4\%$). As anticipated,  the exclusions that derive from the binned-likelihood analysis of the~$m_{4 \ell}$ spectrum are significantly weaker than those that follow from the~$D_S$ distribution. It is also evident from the figure that the size of the assumed systematic uncertainties plays a non-negligible role in the extraction of the  95\%~CL~limits in the~$m_{\phi} \hspace{0.25mm}$--$\hspace{0.25mm} |c_\phi|$ plane, in particular, if the~$m_{4 \ell}$ spectrum is used to discriminate between the BSM signal and the SM background.  We elaborate on this point further  in~Appendix~\ref{app:errorstudy}.  In~this context, we also add that our bounds following from the binned-likelihood analysis of the~$m_{4 \ell}$ distribution agree roughly with the HL-LHC limits presented in~\cite{Goncalves:2017iub,Goncalves:2018pkt}  if one takes into account that these articles have considered the complex Higgs portal~$|H|^2 |\Phi|^2$. A thorough comparison with the latter results is however not possible because a discussion of systematic uncertainties is missing in the works~\cite{Goncalves:2017iub,Goncalves:2018pkt}.  Notice finally that the bounds on~$|c_\phi|$ that follow from our~$D_S$ likelihood-analysis have a non-trivial behaviour for~$m_\phi \lesssim 100 \, {\rm GeV}$. This feature is related to the interference  between the BSM signal and the SM background.  

\section{HE-LHC and FCC analyses}
\label{sec:numHELHCFCC} 

\begin{figure}[!t]
\begin{center}
\includegraphics[width=0.575\textwidth]{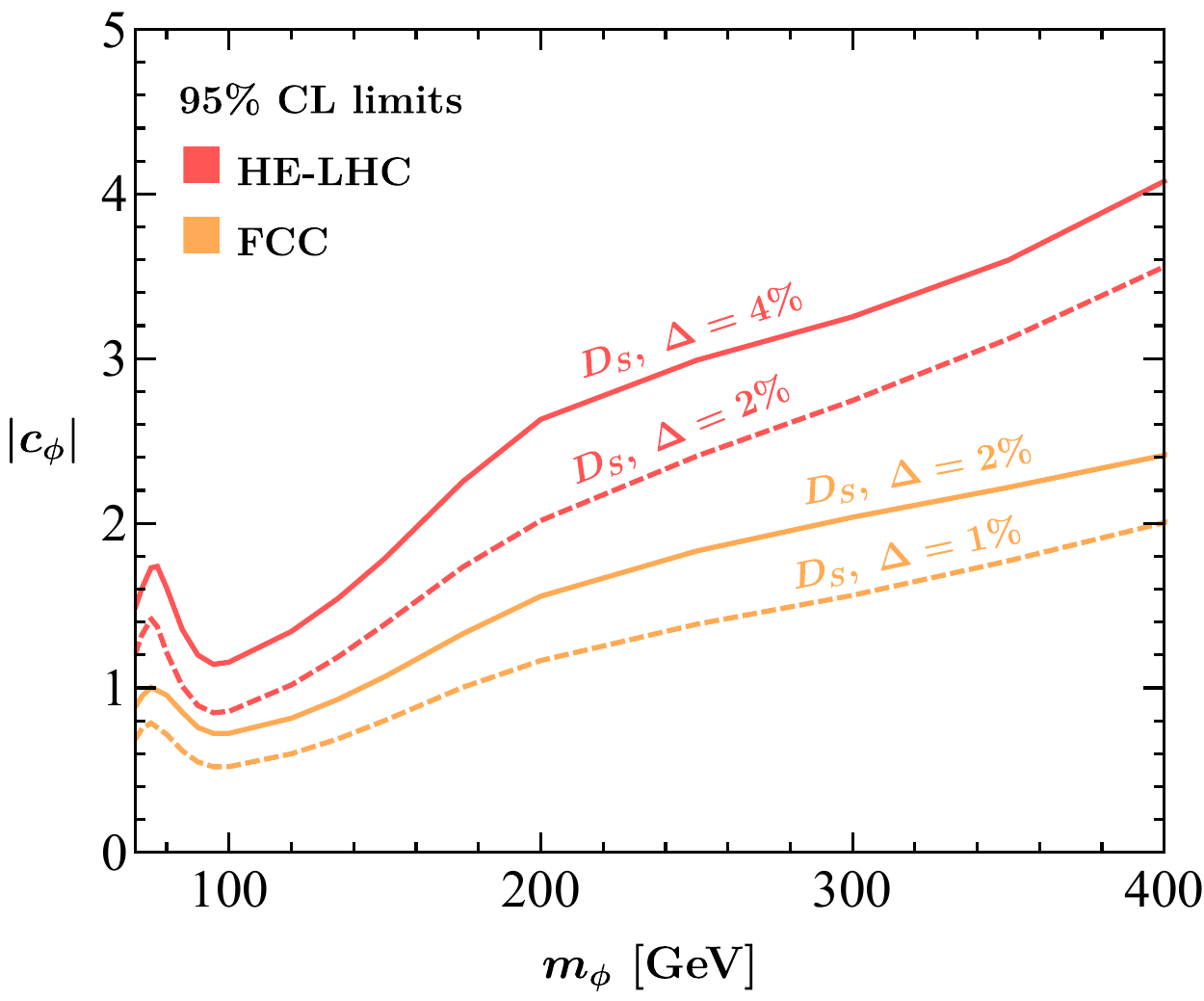} 
\vspace{2mm} 
\caption{\label{fig:HELHCFCCfits}  95\%~CL limits on~$|c_\phi|$  as a function of~$m_\phi$ derived from the binned-likelihood analysis of the ME-based kinematic discriminant~$D_S$. The red and orange exclusions illustrate our HE-LHC and FCC projections, respectively.  The systematic uncertainties that have been assumed to obtain the different bounds are shown next to the lines and vary between~$\Delta = 4 \%$ and~$\Delta = 1 \%$. Further details are given in the main text. }    
\end{center}
\end{figure}

In the following we repeat the numerical analysis performed at the end of the last section  for the HE-LHC and the FCC. In the case of the HE-LHC (FCC) we assume a centre-of-mass energy of~$\sqrt{s} = 27 \, {\rm TeV}$~($\sqrt{s} = 100 \, {\rm TeV}$) and an integrated luminosity of~$15 \, {\rm ab}^{-1}$~($30 \, {\rm ab}^{-1}$). Apart from the~$m_{4\ell}$ window which we enlarge to~$1000 \, {\rm GeV}$~($1500 \, {\rm GeV}$) at the HE-LHC~(FCC), the selection cuts and  detection efficiencies in our HE-LHC and FCC~$pp \to ZZ \to 4 \ell$ analyses  resemble the ones spelled out at the beginning of Section~\ref{sec:numHLLHC}. Possible reductions of the statistical uncertainties due to improvements in the HE-LHC and FCC detectors such as extended pseudorapidity coverages~\cite{Zimmermann:2651305,Benedikt:2651300} are not considered in our numerical analysis. We~also take  the values of the~$K$-factors quoted  in Section~\ref{sec:ME} that have been obtained for LHC collisions to calculate QCD-improved predictions for the kinematic variable~$D_S$ a~la~(\ref{eq:procedure}). In view of the fact that the assumed systematic uncertainties  largely determine the HE-LHC and FCC reach in constraining Higgs portal interactions of the form~(\ref{eq:LHphi}), we believe that these simplifications are fully justified. Moreover,  since we have seen at the end of the last section that the ME-based kinematic discriminant~$D_S$ offers a significantly better sensitivity compared to~$m_{4 \ell}$, we will below only consider the former observable when determining the disfavoured regions in the~$m_{\phi} \hspace{0.25mm}$--$\hspace{0.25mm} |c_\phi|$ plane.  

The HE-LHC and FCC results of our shape fit to the~$D_S$ distribution are displayed in Figure~\ref{fig:HELHCFCCfits}. Like in the case of the HL-LHC we show results assuming different baseline scenarios for the assumed systematic uncertainties. In the case of the HE-LHC we employ~$\Delta = 4 \%$ and~$\Delta = 2 \%$, while in our FCC analysis we use~$\Delta = 2 \%$ and~$\Delta = 1 \%$. These systematic uncertainties can be motivated by noticing that the systematic uncertainties at the HE-LHC should be at least as small as  those expected ultimately  at the HL-LHC and that the FCC has a target precision of~$1.8\%$ for the~$pp \to ZZ \to 4 \ell$ channel~\cite{FCC:2018byv}. Envisaging further theoretical and experimental progress a final systematic uncertainty of~$1\%$ at the~FCC does therefore  not seem inconceivable. From the different curves one again sees that the size of the assumed systematic uncertainties plays a notable role in determining the collider reach. Numerically, we find that  halving the systematic uncertainties at the HE-LHC (FCC) leads to improvements of the 95\%~CL bounds on~$|c_\phi|$ of around~25\% (30\%) at~$m_\phi \simeq 100 \, {\rm GeV}$ and about~20\% (25\%) at~$m_\phi \simeq 250 \, {\rm GeV}$. The gain in statistical power of the FCC compared to the HE-LHC is however also visible from the figure with the FCC bound at~$m_\phi \simeq 250 \, {\rm GeV}$ being better by roughly 25\% than that of the HE-LHC assuming the same systematic uncertainties of~$\Delta = 2\%$. This trend continues at higher values of the real scalar mass reaching up to almost 35\% at~$m_\phi \simeq 400 \, {\rm GeV}$. 

\section{Discussion and outlook}
\label{sec:conclusions} 

\begin{figure}[!t]
\begin{center}
\includegraphics[width=0.575\textwidth]{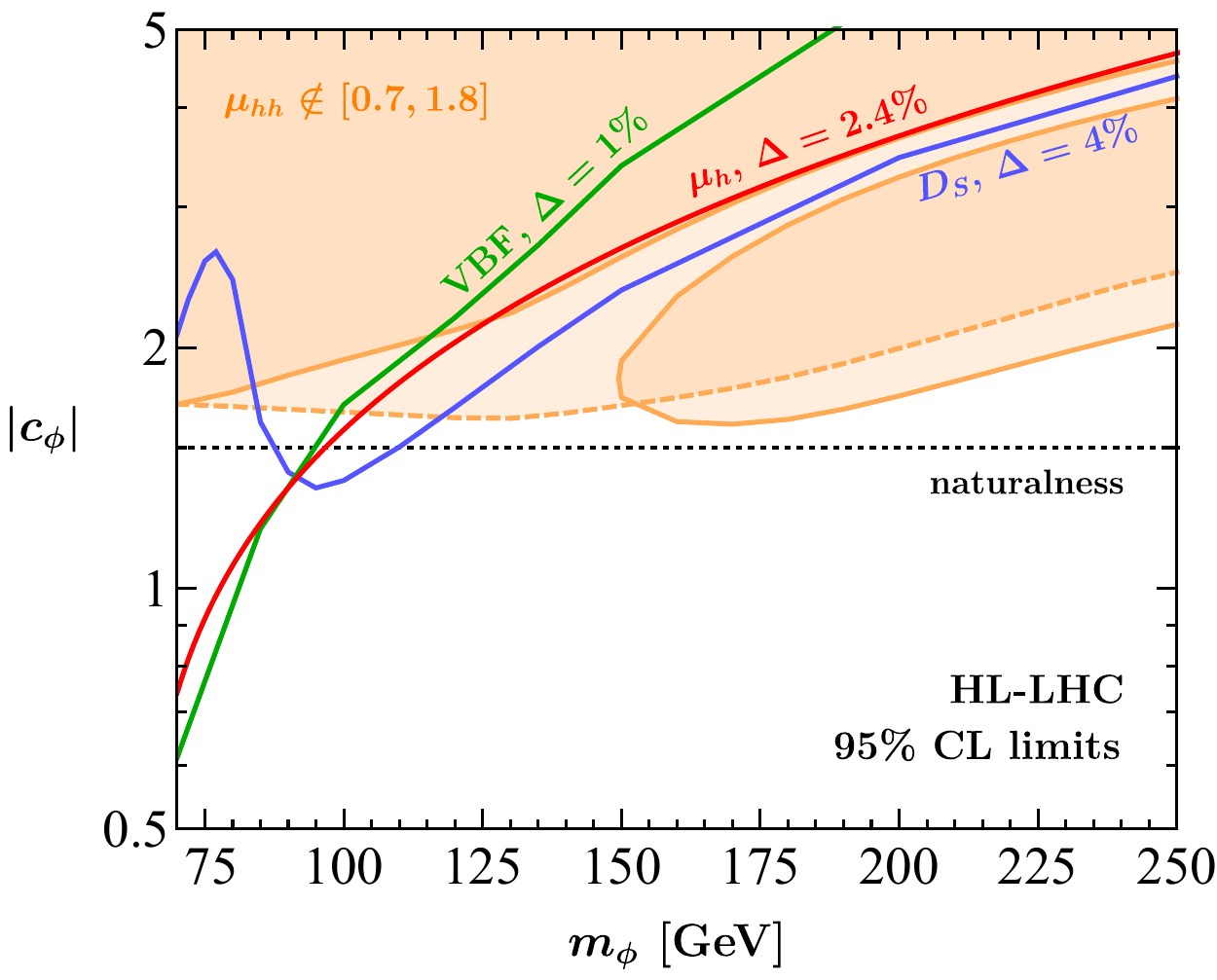} 
\vspace{2mm} 
\caption{\label{fig:HLLHCcomparison} Comparison of the HL-LHC reach of different search strategies in the~$m_{\phi} \hspace{0.25mm}$--$\hspace{0.25mm} |c_\phi|$ plane. The solid blue, solid green and  solid red  line correspond to the  95\%~CL limits that derive from our binned-likelihood analysis of the ME-based kinematic discriminant~$D_S$, the VBF analysis performed in~\cite{Ruhdorfer:2019utl} and  a hypothetical measurement of the global Higgs signal strength~$\mu_h$, respectively. If applicable the assumed systematic uncertainties or accuracies are indicated.  The parameter spaces above the coloured lines are disfavoured. The region bounded by the solid~(dashed) orange line  follows from imposing that the signal strength in double-Higgs production obeys~$\mu_{hh} \notin [0.7,1.8]$  for~$c_\phi > 0$ ($c_\phi < 0$). The dotted black line  corresponds to the bound~$|c_\phi| = \sqrt{3} y_t^2=1.5$ that derives from  naturalness arguments in models of neutral naturalness.  For more details see main text. }    
\end{center}
\end{figure}

In  Figure~\ref{fig:HLLHCcomparison} we compare the HL-LHC reach of different search strategies in the~$m_{\phi} \hspace{0.25mm}$--$\hspace{0.25mm} |c_\phi|$ plane. The solid blue exclusion line corresponds to the 95\%~CL limits that derives from the proposed binned-likelihood analysis of the ME-based kinematic discriminant~$D_S$ assuming a systematic uncertainty of~$\Delta = 4\%$. The solid green line instead indicates the bound obtained in~\cite{Ruhdorfer:2019utl} from a study of off-shell Higgs production in the VBF channel. This analysis assumes a  systematic uncertainty of~$\Delta = 1\%$. At the HL-LHC, measurements of the global Higgs signal strength~$\mu_h$ are expected to reach an accuracy of~$\Delta = 2.4\%$ in the  baseline scenario  S2 for the expected total  systematic uncertainties~\cite{ATL-PHYS-PUB-2018-054}. Utilising the quoted precision together with (\ref{eq:proping}) and (\ref{eq:signalstrengths}) leads at 95\%~CL  to the  solid red line. Another process that is sensitive to Higgs portal interactions of the form~(\ref{eq:LHphi}) is double-Higgs production as previously demonstrated in~\cite{Curtin:2014jma,He:2016sqr,Kanemura:2016lkz,Ruhdorfer:2019utl,Englert:2019eyl}. The 95\%~CL bound~$\kappa_\lambda \in [0.18, 3.6]$ on the modifications~$\kappa_\lambda = \lambda/\lambda_{\rm SM}$ with~$\lambda_{\rm SM} = m_h^2/(2 v^2) \simeq 0.13$  of the trilinear Higgs coupling~as found by the CMS projection~\cite{CMS-PAS-FTR-18-019} implies~$\mu_{hh} \in[ 0.7, 1.8]$ on the  signal strength  in double-Higgs production at the HL-LHC.  By implementing the full one-loop corrections due to~(\ref{eq:LHphi}) into  {\tt MCFM} and imposing the latter bound we obtain the solid and dashed orange lines. Consult Appendix~\ref{app:hhdetails} for further details. Finally, the dashed black  line corresponds to the naturalness bound~$|c_\phi| = \sqrt{3} y_t^2=1.5$ discussed  in~Section~\ref{sec:introduction}. 

\begin{figure}[!t]
\begin{center}
\includegraphics[width=0.575\textwidth]{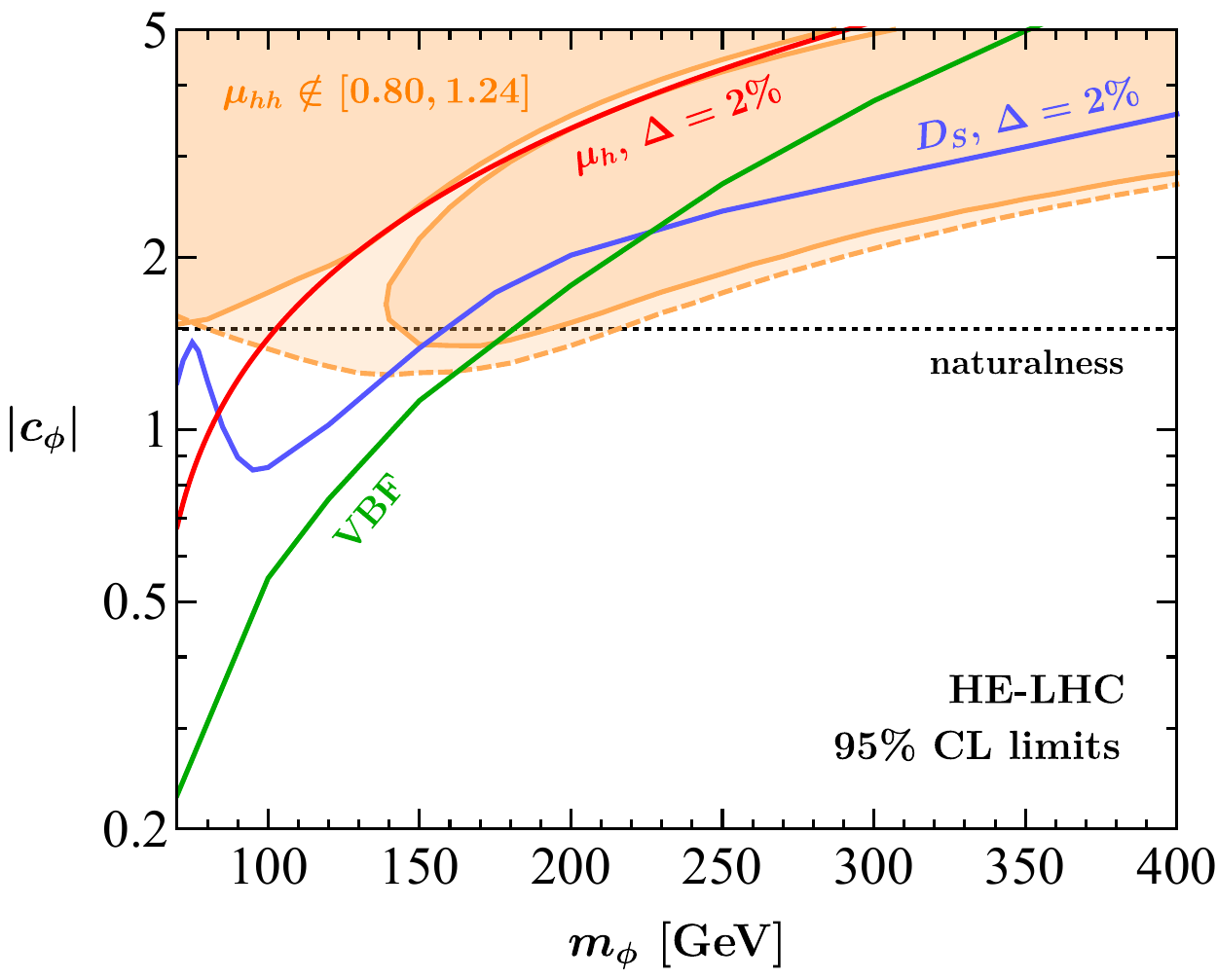} 

\vspace{3mm}

\includegraphics[width=0.575\textwidth]{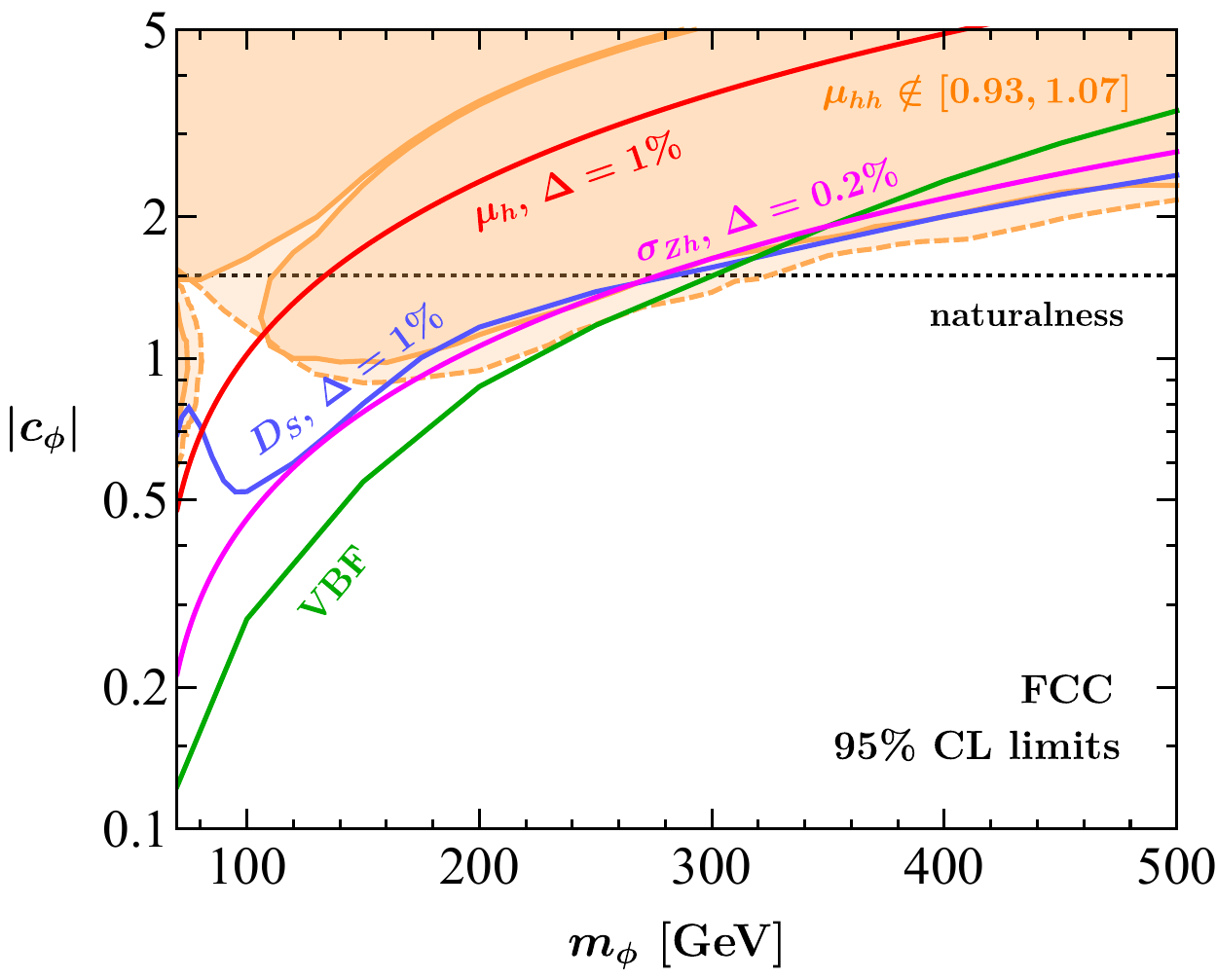} 
\vspace{2mm} 
\caption{\label{fig:HELHCFCCcomparison} Comparison of the HE-LHC~(upper panel) and FCC~(lower panel) reach of different search strategies in the~$m_{\phi} \hspace{0.25mm}$--$|c_\phi|$ plane. Besides the constraints shown in Figure~\ref{fig:HLLHCcomparison}  also the  95\%~CL limit that follows from a precision measurement of the~$Zh$ production cross section~$\sigma_{Zh}$ is displayed in the case of the FCC as a solid magenta line. The colour coding and meaning of the other constraints resembles those in the former figure. Consult the main text for additional explanations.}    
\end{center}
\end{figure}

From Figure~\ref{fig:HLLHCcomparison} it is evident  that for~$m_\phi \lesssim 90 \, {\rm GeV}$ the VBF   and~$\mu_h$ projections provide nominally the best constraints at the HL-LHC. In the case  of~$m_\phi \gtrsim 90 \, {\rm GeV}$, on the other hand, double-Higgs production at the HL-LHC typically allows to set the most stringent constraints on the parameters appearing in~(\ref{eq:LHphi}). Notice also  that the~$D_S$ constraint provides the best  sensitivity for~$90 \, {\rm GeV} \lesssim m_{\phi} \lesssim  120 \, {\rm GeV}$ and stronger constraints than VBF and~$\mu_h$ for~$m_{\phi} \gtrsim 90 \, {\rm GeV}$. The fact that the constraints that stem from double-Higgs production are not symmetric under~$c_\phi \leftrightarrow -c_\phi$ is readily understood by noting that the Higgs portal corrections to the~$gg \to hh$ amplitude involve both terms proportional to~$c_\phi^3$ and~$c_\phi^2$. In fact, integrating out the real scalar~$\phi$ leads to the following one-loop modification of the trilinear Higgs coupling (see for instance~\cite{Ruhdorfer:2019utl,Haisch:2020ahr}):
\beq
\label{eq:hhhmatching}
\kappa_\lambda \simeq 1 + \frac{v^2 c_\phi^2}{12 \pi^2 m_\phi^2}  \left ( \frac{v^2 c_\phi}{m_h^2} - \frac{7}{12} \right ) \,,
\eeq
where the terms in brackets interfere destructively (constructively) for~$c_\phi > 0$ ($c_\phi < 0$). We add that the numerical value of the second term in brackets depends on the definition and the kinematics of the trilinear Higgs vertex~and that the value~in~(\ref{eq:hhhmatching}) is obtained  from the full one-loop form factor~(\ref{eq:renhhhvert}) assuming two on-shell external Higgs bosons. The intricate dependence of the~$gg \to hh$ amplitude on~$m_\phi$ and~$c_\phi$ also leads in the case of~$c_\phi > 0$ to the island of disfavoured parameters starting at~$m_\phi \simeq 145 \, {\rm GeV}$ and~$c_\phi \simeq 1.7$. This point is discussed in more detail in Appendix~\ref{app:hhdetails}. Notice furthermore that all  constraints  shown in  Figure~\ref{fig:HLLHCcomparison}  depend in a non-negligible way on the assumed systematic uncertainties or accuracies. Finally, the VBF limit  only applies if the new degrees of freedom produced in~$h^\ast \to \phi \phi$ are collider stable and thus lead to a missing transverse energy~($E_T^{\rm miss}$) signal at the HL-LHC. In view of these caveats one can conclude that to fully exploit the HL-LHC potential in probing Higgs portal interactions of the form~(\ref{eq:LHphi}) one should consider all direct and indirect probes displayed in Figure~\ref{fig:HLLHCcomparison}. But even in such a case one sees that  at the HL-LHC only theories compatible with the naturalness bound can be explored  if the new particles that cancel the quadratic sensitivity of the Higgs mass are  not heavier than~$m_\phi \simeq 110 \, {\rm GeV}$.  

In the case of the HE-LHC and the FCC   the sensitivity  of the different search strategies to the Higgs portal parameters is  shown in the two panels of Figure~\ref{fig:HELHCFCCcomparison}.  The displayed~$D_S$ constraints assume systematic uncertainties of~$\Delta = 2\%$ and~$\Delta = 1\%$, while the VBF limits taken from~\cite{Ruhdorfer:2019utl} include only statistical uncertainties.  In the case of the global Higgs signal strength~$\mu_h$, we employ~$\Delta = 2\%$ and~$\Delta = 1\%$~\cite{FCC:2018byv}. The 95\%~CL bounds on modifications of the trilinear Higgs coupling at the HE-LHC and the FCC are expected to be~$\kappa_\lambda \in [0.7, 1.3]$ and~$\kappa_\lambda \in [0.9, 1.1]$, respectively. See for example~\cite{Goncalves:2018qas,Bizon:2018syu,Cepeda:2019klc} for detailed discussions.  The~corresponding two-sided limits on the signal strength in double-Higgs production are~$\mu_{hh} \in[ 0.80, 1.24]$ and~$\mu_{hh} \in [0.93, 1.07]$. In addition, we show in the case of the FCC the exclusion that follows from an extraction  of the~$Zh$ cross section~$\sigma_{Zh}$ with an accuracy of~$\Delta = 0.2\%$  as a solid magenta  line. Such a precision measurement should be possible at the~$e^+ e^-$ predecessor of the  FCC running at a centre-of-mass energy of~$\sqrt{s} = 240 \, {\rm GeV}$ with an integrated luminosity of~$5 \, {\rm ab}^{-1}$~\cite{deBlas:2019rxi}. The overall picture observed at the HE-LHC is very similar to that seen at the HL-LHC. Nominal the strongest constraint arises for $m_\phi \lesssim 170 \, {\rm GeV}$ ($m_\phi \gtrsim 170 \, {\rm GeV}$) from VBF  off-shell Higgs (double-Higgs) production,  but the $D_S$ constraint also provides complementary sensitivity in particular for higher values of~$m_\phi$. In the case of the FCC, one furthermore observes that a high precision measurement  of~$\sigma_{Zh}$ can provide additional relevant bounds in the $m_{\phi} \hspace{0.25mm}$--$|c_\phi|$ plane. The combination of all constraints shown in the panels of Figure~\ref{fig:HELHCFCCcomparison} should allow to probe  natural BSM theories  of the form~(\ref{eq:LHstop})  if the new particles that cancel the quadratic sensitivity of the Higgs mass appear below approximately $m_\phi \simeq 200 \, {\rm GeV}$ ($m_\phi \simeq 300 \, {\rm GeV}$) at the HE-LHC~(FCC). 

We add that the potential of CLIC and  a muon collider  in constraining Higgs portal interactions of the form~(\ref{eq:LHphi})  through VBF off-shell Higgs production has been studied in the article~\cite{Ruhdorfer:2019utl}. See also~\cite{Matsumoto:2010bh,Kanemura:2011nm,Chacko:2013lna,Ko:2016xwd} for similar analyses concerning the reach of future lepton colliders. While CLIC  is not expected to improve the FCC bounds  shown in the lower panel of~Figure~\ref{fig:HELHCFCCcomparison} even when running at a centre-of-mass energy of~$\sqrt{s} = 3 \, {\rm TeV}$ and collecting~$3 \, {\rm ab}^{-1}$ of data, a muon collider with~$\sqrt{s} = 6 \, {\rm TeV}$ and~$6 \, {\rm ab}^{-1}$ ($\sqrt{s} = 14 \, {\rm TeV}$ and~$14 \, {\rm ab}^{-1}$) should allow to test natural theories of neutral naturalness up to~$m_\phi \simeq 500 \, {\rm GeV}$ ($m_\phi \simeq 900 \, {\rm GeV}$) thereby exceeding (significantly) the FCC reach.

\acknowledgments
We thank Maximilian~Ruhdorfer, Ennio~Salvioni and Andreas~Weiler for useful discussions, for helpful comments on the manuscript and for providing us with the limits of their VBF analysis~\cite{Ruhdorfer:2019utl}  in electronic form.  The Max~Planck~Computing and Data Facility~(MPCDF) in Garching has been used to carry out the MC simulations related to this work.  Our computations made use of the {\tt Mathematica} packages  {\tt FeynRules}~\cite{Alloul:2013bka}, {\tt FeynArts}~\cite{Hahn:2000kx}, {\tt FormCalc}~\cite{Hahn:1998yk,Hahn:2016ebn} and {\tt Package-X}~\cite{Patel:2015tea}. 

\begin{appendix}

\section{Systematic uncertainties}
\label{app:errorstudy}

\begin{figure}[!t]
\begin{center}
\includegraphics[width=0.575\textwidth]{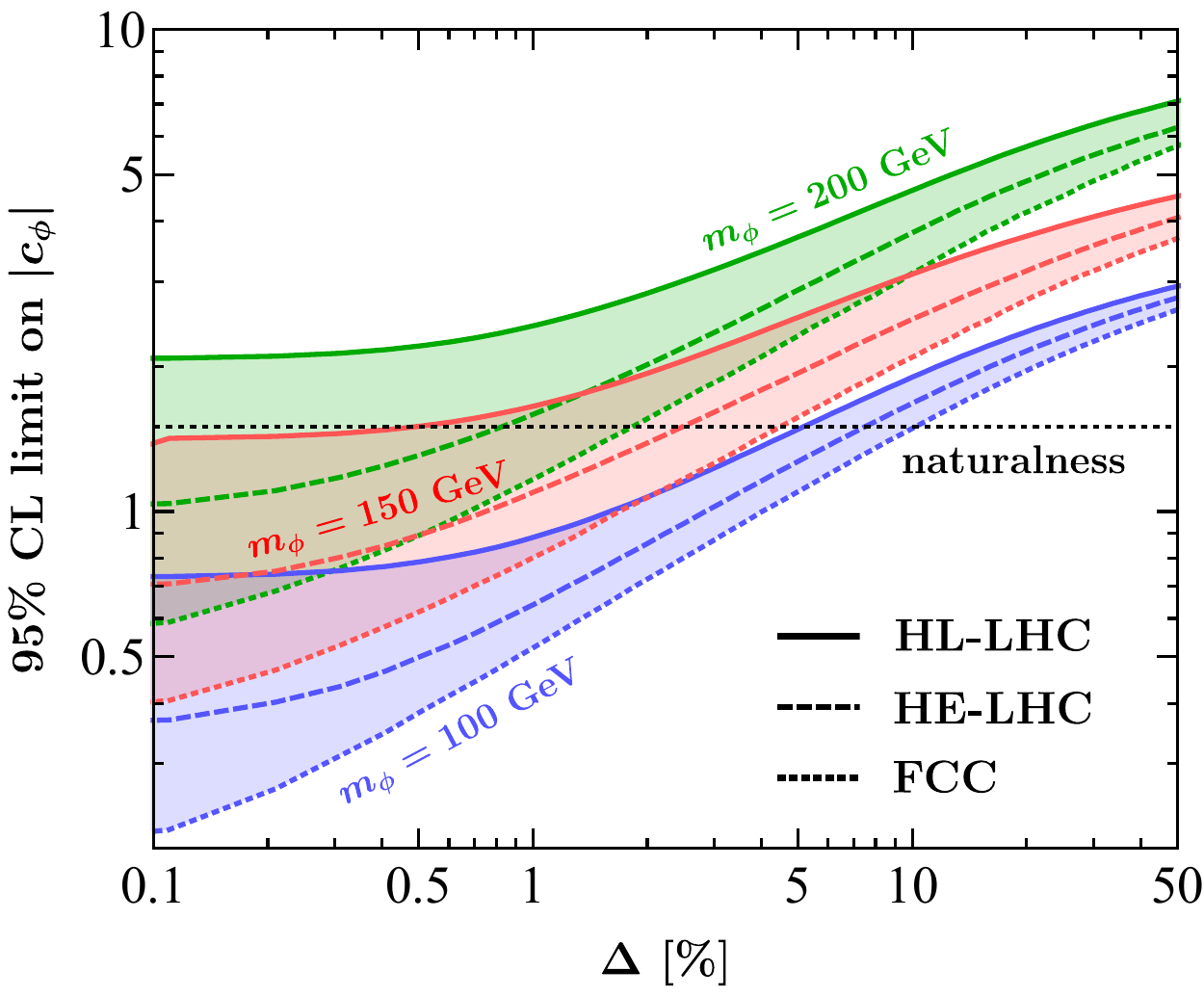}  
\vspace{2mm} 
\caption{\label{fig:syst_errs}  95\%~CL limits on~$|c_\phi|$ derived from the binned-likelihood analysis of the~$D_S$ spectra as a function of the assumed systematic uncertainty $\Delta$. The bounds for the HL-LHC~(solid lines), the HE-LHC~(dashed lines) and the FCC~(dotted lines) are displayed for the three benchmark values $m_{\phi} = 100 \, {\rm GeV}$~(blue), $m_{\phi} = 150  \, {\rm GeV}$~(red) and $m_{\phi} = 200 \, {\rm GeV}$~(green) of the scalar mass. The dotted black line  corresponds to the condition~$|c_\phi| = \sqrt{3} y_t^2=1.5$ that derives from  naturalness arguments in models of neutral naturalness. See main text for additional details. }
\end{center}
\end{figure}

In this appendix we discuss in more detail the prospects of the proposed binned-likelihood analyses of the $D_S$~spectra for the HL-LHC, the HE-LHC and the FCC. In particular, we examine how different assumptions on the systematic uncertainties affect the resulting constraints on the parameter space of the  Higgs portal model~(\ref{eq:LHphi}). In Figure \ref{fig:syst_errs}, we show the projected  95\%~CL limits on~$|c_\phi|$ derived from our $D_S$ analysis as a function of the assumed systematic uncertainty $\Delta$ for the three aforementioned colliders. The presented limits are obtained using the benchmark numerical values for the scalar masses indicated in the figure that vary between $100 \, {\rm  GeV}  \leq m_{\phi}\leq 200 \, {\rm  GeV}$.  

Figure~\ref{fig:syst_errs} further illustrates the point already made in Sections~\ref{sec:numHLLHC} and~\ref{sec:numHELHCFCC}, that the assumptions on the systematic uncertainties~$\Delta$ play a crucial role in  constraining the  $m_{\phi} \hspace{0.25mm}$--$\hspace{0.25mm} |c_\phi|$ parameter space by using the $D_S$ distribution as a kinematic discriminant. In particular, one observes that the enhanced statistical power provided by the HE-LHC and the~FCC, which results from the increased centre-of-mass energy and integrated luminosity of these machines  compared to the HL-LHC, can  only be fully exploited if systematic uncertainties are under control. For instance,  in the case of $m_{\phi} = 100 \, {\rm GeV}$ the sensitivity gain between the HL-LHC and the FCC  is around 17\%  for $\Delta = 20\%$, while for $\Delta = 1\%$ the improvement  amounts  to about 41\%. Similar numbers of approximately 26\% and 51\% are found for $m_{\phi} = 150 \, {\rm GeV}$ and $m_{\phi} = 200 \, {\rm GeV}$, implying that the gain in sensitivity  between different colliders is to first approximation mass-independent for the low values~of~$m_\phi$ considered in the figure.

\section{\boldmath Details of the double-Higgs calculation}
\label{app:hhdetails}

At the one-loop level the~$gg \to hh$ process receives contributions from virtual~$\phi$ exchange in propagator and vertex diagrams as well as counterterm contributions associated to wave function, mass and tadpole renormalisation (see~\cite{He:2016sqr, Englert:2019eyl} for details). In the on-shell scheme the combined corrections involving the Wilson coefficient~$c_\phi$ can be written as a finite shift:
\beq
\lambda_{\rm SM} \to \lambda_{\rm SM} \, \big  [ 1 + \delta (\hat s) \big ] \,.  
\eeq
Here~$\lambda_{\rm SM} = m_h^2/(2 \hspace{0.125mm} v^2)$ is the  tree-level  expression for the trilinear Higgs coupling in the SM and the~$\hat s$-dependent form factor is given by 
\beq \label{eq:renhhhvert}
\begin{split}
\delta (\hat s) & =   -\frac{ v^2  \hspace{0.25mm} c_\phi ^2 }{24 \hspace{0.125mm}\pi^2 \hspace{0.125mm} m_h^2} \hspace{0.25mm} \left ( 1 + \frac{3 \hspace{0.25mm} m_h^2}{\hat{s}-m_h^2} \right ) \Big[ B_0 \! \left (\hat s,m_\phi^2,m_\phi^2 \right ) -B_0 \! \left (m_h^2,m_\phi^2,m_\phi^2 \right ) \Big] \\[2mm]
& \phantom{xx} - \frac{ v^4  \hspace{0.25mm} c_\phi ^3 }{6 \hspace{0.125mm} \pi^2 \hspace{0.125mm} m_h^2} \hspace{0.5mm}  C_0 \! \left (m_h^2,m_h^2,\hat s,m_\phi^2,m_\phi^2,m_\phi^2 \right ) - \frac{v^2  \hspace{0.25mm} c_\phi ^2 }{8 \hspace{0.125mm} \pi^2} \left . \frac{d}{d \hat s}  B_0 \! \left (\hat s,m_\phi^2,m_\phi^2 \right ) \right |_{\hat s = m_h^2}  \, ,
\end{split}
\eeq
with the~$A_0$,~$B_0$ and~$C_0$ functions are one-,  two-, and three-point Passarino-Veltman scalar integrals defined as in~\cite{Hahn:1998yk,Hahn:2016ebn}. Our result (\ref{eq:renhhhvert}) agrees with~\cite{He:2016sqr, Englert:2019eyl}, after fixing a sign error in~(12) of \cite{Englert:2019eyl}. Notice that after integrating out the scalar field~$\phi$ by expanding the on-shell form factor~$\delta (2 \hspace{0.25mm} m_h^2)$ up to the first power in~$m_h^2 / m_{\phi}^2$, one recovers the approximate correction for~$\kappa_\lambda$ as given  in~(\ref{eq:hhhmatching}).

\begin{figure}[!t]
\begin{center}
\includegraphics[width=0.575\textwidth]{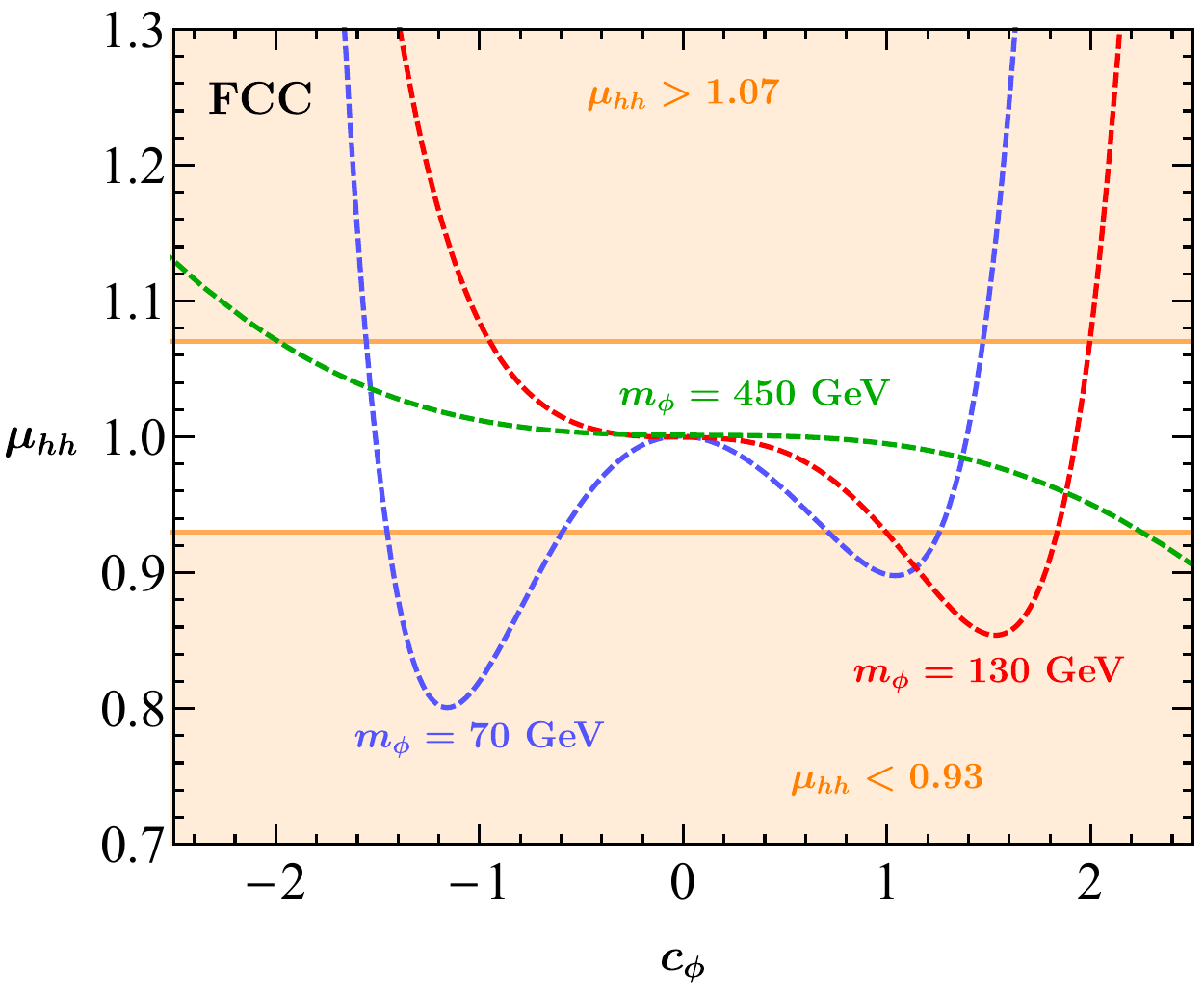} 
\vspace{2mm} 
\caption{\label{fig:muhh_FCC} The signal strength for double-Higgs production~($\mu_{hh}$) at the FCC as a function of the Wilson coefficient~$c_{\phi}$ for three values of the scalar mass:~$m_{\phi} = 70 \, {\rm GeV}$~(dashed blue),~$m_{\phi} = 130 \, {\rm GeV}$~(dashed red) and~$m_{\phi} = 450 \, {\rm GeV}$~(dashed green). The regions excluded by the projected experimental constraint~$\mu_{hh} \in [0.93, 1.07]$ are shown in orange. For further explanations see main text. }
\end{center}
\end{figure}

To obtain predictions for double-Higgs production we have implemented the analytic results~(\ref{eq:renhhhvert})  at the amplitude level into {\tt MCFM}. We then perform sensitivity scans in  the parameters~$c_{\phi}$ and~$m_{\phi}$, using the setup discussed at the beginning of Section~\ref{sec:numHLLHC}, but fixing the renormalisation and factorisation scales~$\mu_R$ and~$\mu_F$  to the value~$2 \hspace{0.125mm} m_h$. In Figure~\ref{fig:muhh_FCC}~we show results for the  signal strength~$\mu_{hh}$ in double-Higgs production for three different values of~$m_\phi$ as a function of~$c_\phi$. The displayed curves correspond to the results obtained at the~FCC. Two feature of the shown predictions deserve some comments. First, due to the~$c_\phi^3$ and~$c_\phi^2$ dependence of~(\ref{eq:renhhhvert}) the signal strengths~$\mu_{hh}$ are not symmetric under~$c_\phi \leftrightarrow -c_\phi$. Second, the functional form of~$\mu_{hh}$ depends also sensitively on the mass~$m_\phi$. For low~$\phi$ masses as illustrated by the choice~$m_\phi = 70 \, {\rm GeV}$ in the figure, the signal strength~$\mu_{hh}$ has two minima, one at around~$c_\phi \simeq -1.1$ and another one at~$c_\phi \simeq 1.0$. This feature leads to the orange exclusions in the lower plot in Figure~\ref{fig:HELHCFCCcomparison} at~$|c_\phi | \simeq 1$. For larger values of~$m_\phi$  the signal strengths~$\mu_{hh}$ have  instead only a single minimum at positive values of~$c_\phi$. Notice that if the value of~$\mu_{hh}$ at this minimum is incompatible with the experimental allowed range, such as happens to be the case for example for~$m_\phi = 130 \, {\rm GeV}$ at the FCC, increasing/decreasing  the value of~$c_\phi$ will always  result in~$\mu_{hh}$ values that are  consistent with experiment. This feature leads to the orange exclusions shown in the plots of Figures~\ref{fig:HLLHCcomparison} and~\ref{fig:HELHCFCCcomparison} that are relevant for   $c_\phi > 0$ and separated by a funnel of viable solutions. 

\end{appendix}



\providecommand{\href}[2]{#2}\begingroup\raggedright\endgroup

\end{document}